\documentclass[aps,prd,twocolumn,preprintnumbers,amsmath,amssymb,amsfonts,nofootinbib,superscriptaddress,altaffilletter]{revtex4-2}

\usepackage{graphicx}
\usepackage[unicode]{hyperref}
\hypersetup{
  bookmarksopen=true
}
\usepackage{amssymb}
\usepackage{amsmath}
\usepackage{braket}
\usepackage{mathtools, lipsum}
\usepackage{color}
\usepackage{supertabular}
\usepackage{dcolumn}
\usepackage{multirow}
\usepackage{upgreek}
\usepackage{xcolor}
\usepackage{booktabs}
\usepackage{natbib}
\usepackage{float}
\usepackage{orcidlink}
\usepackage{comment}
\usepackage{xspace}

\newif\ifshowfigs
\showfigstrue

\usepackage{soul}

\newcommand{\useMSU}[1]{}
\def\indD{IndIGO-D\xspace}

\begin{document}

\title{Prospects for Direct Detection of Ultralight Dark Matter Candidates in the deci-Hz Band with IndIGO-D}

\author{Md. Emanuel Hoque\,\orcidlink{0009-0002-8488-8758}}
\email{emanuel.hoque@saha.ac.in}
\affiliation{Saha Institute of Nuclear Physics, 1/AF Bidhannagar, Kolkata-700064, India}
\affiliation{Homi Bhabha National Institute, Anushakti Nagar, Mumbai 400094, India}

\author{Andrew L. Miller,\orcidlink{0000-0002-4890-7627}}
\email{andrew.miller.ligo@ucas.ac.cn}
\affiliation{International Centre for Theoretical Physics Asia-Pacific (ICTP-AP), University of Chinese Academy of Sciences (UCAS), Beijing 100190, China.}
\affiliation{Taiji Laboratory for Gravitational Wave Universe, University of Chinese Academy of Sciences, 100049 Beijing, China}

\author{Arunava Mukherjee\,\orcidlink{0000-0003-1274-5846}}
\email{arunava.mukherjee@saha.ac.in}
\affiliation{Saha Institute of Nuclear Physics, 1/AF Bidhannagar, Kolkata-700064, India}
\affiliation{Homi Bhabha National Institute, Anushakti Nagar, Mumbai 400094, India}
\date{\today}

\begin{abstract}
We investigate the sensitivity of IndIGO-D, a proposed space-based decihertz gravitational-wave interferometer, to different classes of ultralight dark matter. IndIGO-D will probe the $\sim0.01$--$10~\mathrm{Hz}$ frequency band between those accessible to current ground- and future space-based gravitational-wave interferometers, providing access to ultralight dark-matter masses beyond the reach of existing instruments. We consider two complementary signatures: interferometric displacements induced by coherently oscillating scalar (dilaton), vector (dark-photon, $U(1)_B$ and $U(1)_{B-L}$ gauge groups), and tensor fields with masses $m_{\rm DM}\sim10^{-17}$--$10^{-14}~\mathrm{eV}$; and changes in laser polarization induced by pseudoscalar axion dark matter at higher masses, around $m_a\sim10^{-12}~\mathrm{eV}$. For the dilatons, dark photons and tensors, we compute the expected sensitivities using two pipelines -- cross-correlation and BSD excess-power -- assuming L-shaped and triangular interferometer layouts and three representative noise power spectral densities (S1, S2, S3), each for two years of continuous observation. We find that the projected sensitivities agree to within a factor of order unity across the search pipelines and interferometer geometries. In particular, we show that IndIGO-D could open previously unconstrained coupling parameter space for vector and tensor dark matter across $m_{\rm DM}\sim10^{-16}$--$10^{-14}~\mathrm{eV}$, bridging the sensitivity of space- and ground-based experiments. For axion dark matter, we demonstrate that a complementary detection for laser light polarization shifts, limited primarily by photon shot noise, could probe the axion-photon coupling $g_{a\gamma}$ at masses around $m_a\sim10^{-12}~\mathrm{eV}$ at a level potentially better than that of other future experiments without degrading sensitivity to gravitational waves.
\end{abstract}

\maketitle

\section{Introduction}\label{sec:introduction}

Despite extensive theoretical and observational efforts, the fundamental nature of dark matter (DM) remains one of the greatest unsolved mysteries in physics~\cite{Bertone_2018}. Ultralight particles with masses far below the eV scale have been widely proposed~\cite{PecceiQuinn_PRL1977, Weinberg_PRL1978, Wilczek_PRL1978, Dubovsky_PRL2005, Hassan_JHEP2012, Stadnik_PRL2015, Babichev_PRD2016, Hees_PRD2018, AGRAWAL_PRB2020, Ferreira_2021} as potential candidates for DM. Direct searches for ultralight DM (ULDM) using various ground- and space-based gravitational-wave (GW) interferometers have recently been carried out~\cite{Pierce_2018, Nagano_PRL2019, Armaleo_2021, Manita_2023, Yu_etal_prd2023, Yao_and_Tang_prd2024, Yao_etal_prd2025}, since such fields behave as coherent, nearly monochromatic classical waves in the Galactic halo, including our local solar neighborhood. 

A defining feature of ULDM is its enormous phase-space occupation number in the Galactic halo. As a result, these candidates behave not as individual particles but as coherently oscillating classical fields, with oscillation frequencies set by their masses~\cite{Preskill:1982cy,Abbott:1982af,Dine:1982ah}. This coherence endows ULDM with distinctive experimental signatures: narrow-band, nearly monochromatic signals that persist over macroscopic time and length scales~\cite{Miller_2026}. The occupation number and the associated coherent wave frequency are outlined in~\citet{Miller_prd2021}, 

\begin{equation}
\begin{split}
N_0
= \lambda_{\rm dB}^3 \frac{\rho_{\mathrm{DM}}}{m_{\mathrm{DM}}}
= \left( \frac{2\pi}{m_{\mathrm{DM}} v_0} \right)^3
\frac{\rho_{\mathrm{DM}}}{m_{\mathrm{DM}}}
\\ \simeq 1.69 \times 10^{54}
\left( \frac{10^{-12}\,\mathrm{eV}}{m_{\mathrm{DM}}} \right)^4,
\label{eqn:N0_occpn_num}
\end{split}
\end{equation}

\begin{equation}
f_0 = \frac{m_{\mathrm{DM}}}{2\pi}
\simeq 241~\mathrm{Hz} \left( \frac{m_{\mathrm{DM}}}{10^{-12}\,\mathrm{eV}} \right),
\label{eqn:f0_freq}
\end{equation}
where, $m_{\mathrm{DM}}$ is the DM particle mass, $\lambda_{\mathrm{dB}} = 2\pi/(m_{\mathrm{DM}} v_0)$ is its de Broglie wavelength, $v_0 \approx 220~\mathrm{km/s}$ is the local virial velocity~\cite{Smith_2007}, and $\rho_{\mathrm{DM}} = 0.4~\mathrm{GeV/cm^{3}}$ is the local DM density in the Galactic halo~\cite{Salucci_AA2010, de_Salas_2021}. The large occupation number in Eq.~\eqref{eqn:N0_occpn_num} makes the field amplitude effectively classical, while Eq.~\eqref{eqn:f0_freq} establishes a direct correspondence between the DM mass and a measurable oscillation frequency. A laser interferometer immersed in this oscillating field acquires a small, coherent, narrow-band response, making such precision instruments natural detectors for ultralight DM candidates.

However, a GW interferometer is sensitive only to a limited mass range determined by its instrumental, material, and design specifications. Ground-based detectors---the currently operating second-generation Advanced-LIGO, Advanced-Virgo, and KAGRA (LVK)~\cite{LVK_2015, Virgo_2014, Kagra_2020} and future third-generation detectors, e.g., the Einstein Telescope (ET)~\cite{ET_2026} and Cosmic Explorer (CE)~\cite{CE_2023}---target the high-frequency range, while proposed space-based missions like LISA~\cite{LISA_2017} and Taiji~\cite{Taiji_2023} cover the low-frequency band. In comparison, a new space mission, \indD---a conceptualized space-based GW interferometer---aims to probe the deci-Hz ($\sim 0.01$--$10~\mathrm{Hz}$) frequency band to fill the gap between these two regimes.

While the primary goal of these interferometers across different frequency ranges is to detect GWs from astrophysical and cosmological sources, most of them can also be repurposed for detecting different classes of DM candidates. The frequency-mass relation described in Eq.~\eqref{eqn:f0_freq} also identifies specific detectors that are sensitive to certain ranges of particle mass. Ground-based interferometers of the LVK network operate above $\sim\!10~\mathrm{Hz}$, probing masses $m_{\mathrm{DM}} \gtrsim 4\times 10^{-14}~\mathrm{eV}$ \cite{LIGOScientific:2025ttj}, while the space-based LISA mission will reach down to the millihertz band, accessing $m_{\mathrm{DM}} \lesssim 10^{-17}~\mathrm{eV}$. Between these two ranges lies the decihertz band ($0.01$--$10~\mathrm{Hz}$), corresponding to $m_{\mathrm{DM}} \sim 10^{-17}$--$10^{-14}~\mathrm{eV}$. This band is of particular interest for GW astronomy to capture the early inspirals of compact binary coalescences~\cite{LISA-TAIJI}, which has led to the proposal of two missions on the Moon---LGWA~\cite{LGWA_2025} and LILA \cite{Jani:2025uaz}---and two space-based instruments---DECIGO~\cite{DECIGO}, and \indD~\cite{Sharma_2026}. However, as we will illustrate here, these missions, in particular \indD, can also serve as direct DM detectors owing to their exquisite sensitivities to small displacements. A complementary study on probing the Compact Binary Coalescence (CBC) science case for the same mission concept was presented in~\cite{Sharma_2026}. Moreover, such detectors could be sensitive to supernovae at low frequencies~\cite{Choi:2024irp}. 

In parallel with proposed space and lunar GW detectors, long-baseline strontium atom interferometers~\cite{Graham_etal_PRL2013}—such as AION~\cite{Badurina_2020}, AEDGE~\cite{El_Neaj_2020}, and MAGIS-100~\cite{Abe_2021}—alongside terrestrial demonstrators like MIGA~\cite{Canuel_2018} and ZAIGA~\cite{Zhan_2019}, provide a dedicated quantum path for ULDM searches in the deci-Hz band. By measuring time-varying atomic transition frequencies and differential phase shifts, these long-baseline quantum networks target scalar and vector ULDM couplings across the $10^{-19}\text{ to }10^{-12}\text{ eV}$ mass range. 

In this paper, we assess the sensitivities of \indD to different ULDM candidates and place them in the context of DM probes with existing and planned GW instruments (LVK, CE, LISA, and DECIGO), as well as with existing bounds from the E\"ot-Wash~\cite{EotWash_2008} and MICROSCOPE~\cite{MICROSCOPE_2022} experiments. We consider four representative classes of ULDM distinguished by the nature of their coupling to the Standard Model sector: (i) scalar (dilaton) DM, which modulates fundamental constants~\cite{Stadnik_PRL2015}; (ii) vector (dark-photon) DM under the $U(1)_B$ and $U(1)_{B-L}$ gauge groups~\cite{AGRAWAL_PRB2020}; (iii) tensor DM coupling through a Yukawa-type interaction~\cite{Dubovsky_PRL2005, Babichev_PRD2016, Hassan_JHEP2012}; and (iv) pseudoscalar (axion) DM coupling to photons~\cite{PecceiQuinn_PRL1977, Weinberg_PRL1978, Wilczek_PRL1978}.

\section{The \indD detector}\label{sec:detector}

\subsection{Configuration}

\indD is a conceptual space-based GW interferometer proposed under the Indian Initiative in GW Observations (IndIGO) to operate in the decihertz (dHz) band. The mission concept, together with its orbital design, antenna response, and compact-binary science case, is described in detail in a companion paper by~\citet{Sharma_2026}; here we summarize only our findings relevant to the direct detection of different classes of ULDM.

The \indD mission consists of three spacecraft flying in a heliocentric orbit with $1000~\mathrm{km}$ baselines and a vertex spacecraft serving as the laser source. Two interferometer shapes for the constellation of spacecraft are currently under consideration. In the baseline configuration, as adopted in the companion paper~\cite{Sharma_2026}, the two arms are orthogonal and meet at the common vertex, forming a single L-shaped Michelson interferometer—a space-based analogue of a terrestrial detector. In the alternative configuration, the spacecraft are arranged with an inter-arm opening angle of $\pi/3$, which provides two quasi-independent data channels analogous to the $A$ and $E$ channels of LISA~\cite{Romano_2017}. These two constellation shapes call for different DM readout strategies. We find that they yield coupling sensitivities that agree within an order of magnitude; therefore, the projected sensitivities are robust against the design choice for \indD.

The orbital configuration is chosen following the stable-formation principle for LISA-like geometries~\citep{Dhurandhar_2005} and adapted to the L-shaped geometry outlined by~\citet{Sharma_2026}. Under this configuration, the inter-spacecraft separation remains fixed to first order in the small parameter $\alpha = \ell/(2R) \approx 3.3\times10^{-6}$, where $\ell$ is the arm length and $R = 1~\mathrm{AU}$ is the orbital radius of the vertex spacecraft. The residual arm-length flexing is at the level of a few parts per million (ppm) and occurs on a monthly timescale, introducing only $\mathcal{O}(10^{-6})$ fractional corrections to the interferometer response~\cite{Sharma_2026}. For the purposes of this work, the arms may therefore be treated as rigid and of fixed, equal length.\footnote{The slow flexing of the baseline at the few-ppm level over the orbital period is neglected.}

\subsection{Detector Noise Sensitivity Model}

\begin{figure*}[hbt!]
    \centering
    \includegraphics[clip,width=\textwidth]{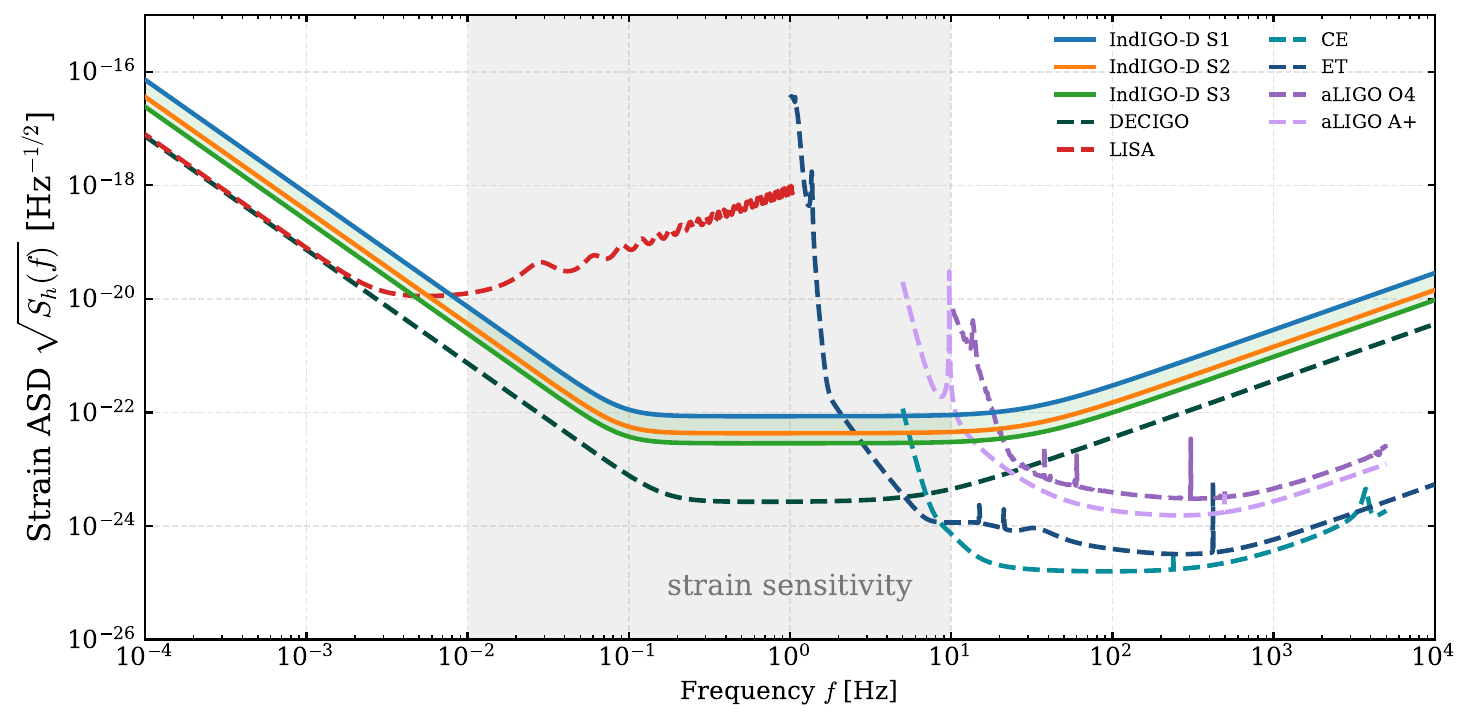} \\
    \caption{The noise amplitude spectral density (ASD) curves for the strain sensitivity of \indD{} are shown for three scenarios (S1, S2, and S3), alongside those of LISA, DECIGO, Advanced LIGO, ET, and CE across a wide frequency range. The light-gray shaded region highlights the targeted sensitivity band for the \indD{} detector.}
    \label{fig:strain_asd}
\end{figure*}
 
We adopt the fiducial one-sided strain noise power spectral density (PSD) of \indD{}~\cite{IndIGO-D_sens},
\begin{equation}
S_n(f) = A_{\mathrm{acc}} \left( \frac{f}{f_l} \right)^{-4}
        + A_{\mathrm{shot}} \left[ 1 + \left( \frac{f}{f_u} \right)^{2} \right],
\label{eqn:psd}
\end{equation}
where the first term captures low-frequency acceleration noise and the second term captures high-frequency shot noise, with corner frequencies $f_l = 0.1~\mathrm{Hz}$ and $f_u = 30~\mathrm{Hz}$ bounding the putative detector bandwidth. A representative realization with $1000~\mathrm{km}$ arms and a $\sim\!10~\mathrm{W}$, $\sim\!1~\mu\mathrm{m}$ wavelength laser is expected to achieve this level of sensitivity.

Since the final design sensitivity remains to be decided, we bracket the projected sensitivity of ULDM searches using three representative GW detector noise scenarios, denoted by S1, S2, and S3. These correspond to three different PSDs with progressively smaller acceleration and shot noise amplitudes $(A_{\mathrm{acc}}, A_{\mathrm{shot}})$ (see Eq.~\eqref{eqn:psd}). The three scenarios are listed in Table~\ref{tab:psd} and span roughly an order of magnitude in strain PSD, illustrating how the sensitivity scales with achievable instrument performance. To place these choices in context, the resulting strain amplitude spectral densities for S1, S2, and S3 are compared against the projected strain PSDs of LISA, DECIGO, Advanced LIGO O4, Advanced LIGO A$^{+}$, Einstein Telescope (ET), and Cosmic Explorer (CE) in Fig.~\ref{fig:strain_asd}.

\begin{table}[htbp]
\caption{\label{tab:psd}Acceleration and shot-noise amplitudes of the three representative \indD detector strain noise sensitivity curves (S1, S2, and S3) used in this work, for the PSD of Eq.~\eqref{eqn:psd} with $f_l = 0.1~\mathrm{Hz}$ and $f_u = 30~\mathrm{Hz}$.}
\renewcommand{\arraystretch}{1.3} 
\begin{ruledtabular}
\begin{tabular}{ccc}
Scenario & $A_{\mathrm{acc}}$ ($\mathrm{Hz}^{-1}$) & $A_{\mathrm{shot}}$ ($\mathrm{Hz}^{-1}$) \\
\hline
S1 & $52.0\times10^{-46}$ & $73.32\times10^{-46}$ \\
S2 & $13.0\times10^{-46}$ & $18.47\times10^{-46}$ \\
S3 & $5.81\times10^{-46}$ &  $8.19\times10^{-46}$ \\
\end{tabular}
\end{ruledtabular}
\end{table}

\section{Cold, Ultralight Dark Matter}\label{sec:uldm}

Cold ULDM interacts with Standard Model particles in various model-dependent ways. Despite this, the enormous number of particles within a given volume motivates treating ULDM as a classical oscillating field, regardless of its specific interactions.

\subsection{Dilaton Scalar Dark Matter}

ULDM could have originated via the \emph{vacuum misalignment mechanism} in the early Universe, manifesting today as a \emph{coherently oscillating scalar field}~\cite{Stadnik_PRL2015}. The coupling of such a field to Standard-Model particles would induce oscillations in fundamental constants, such as the electron rest mass and the fine-structure constant, leading to periodic expansion and contraction of solids. In the presence of spatial gradients of the field, macroscopic accelerations of objects can also occur. 

\begin{figure*}[hbt!]
    \centering
    \includegraphics[clip,width=\textwidth]{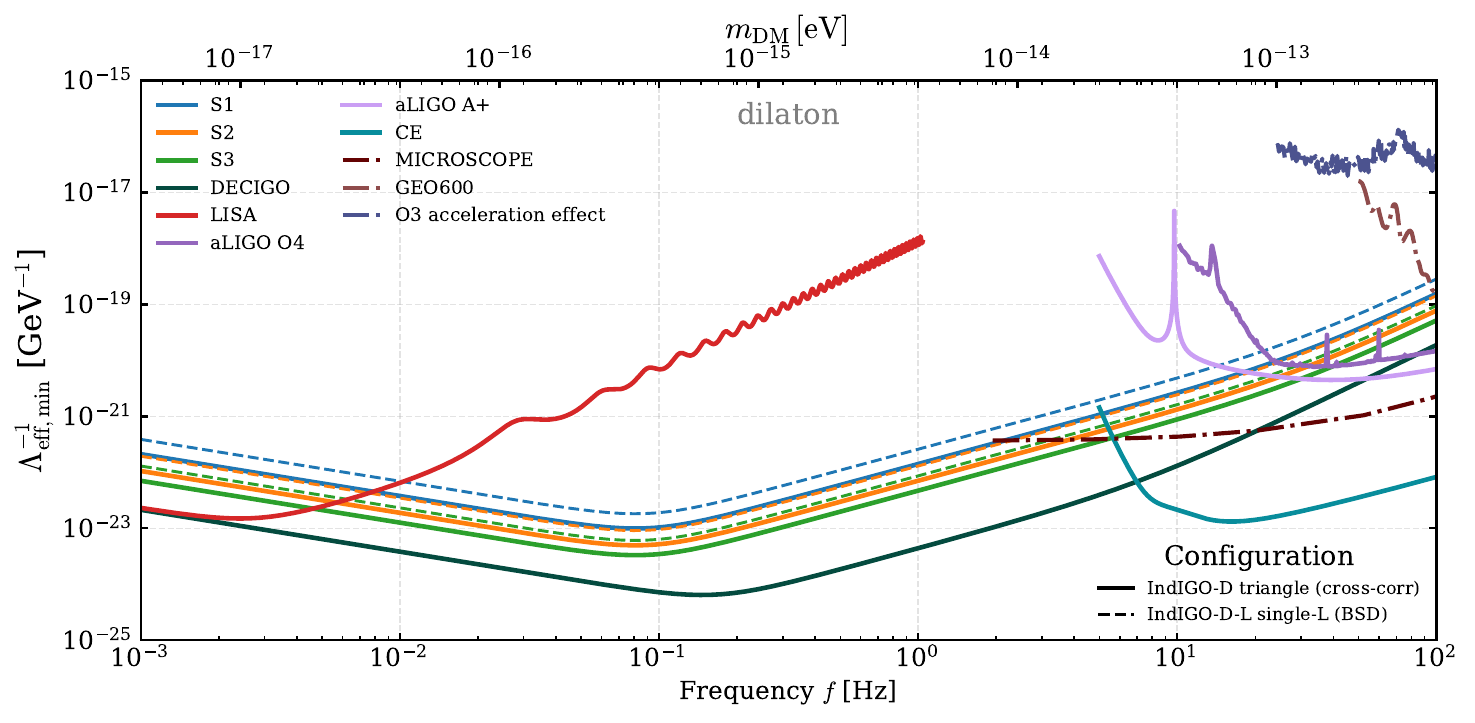} \\ 
    \caption{Projected sensitivity to the effective DM coupling constant $\Lambda^{-1}_{\mathrm{eff}, \mathrm{min}} \equiv (\frac{1}{\Lambda_{\gamma}} + \frac{1}{\Lambda_{e}})_{\mathrm{min}}$ as a function of the DM mass $m_{\rm DM}$, assuming a continuous observation time of 2 years. Sensitivity curves are shown for three representative proposed deciHz GW interferometers (S1, S2, and S3), each with two different configurations: (i) triangular constellation (with solid lines) and (ii) single L-shaped constellation (with dashed lines), as well as for Advanced-LIGO O4, Advanced-LIGO A+, DECIGO, and LISA. Existing experimental constraints from MICROSCOPE~\cite{MICROSCOPE_2022}, GEO600~\cite{Vermeulen_2021}, and the O3 acceleration-effect searches~\cite{O3_DM_2022} are also displayed for comparison. The sensitivities are computed for a detection threshold corresponding to $\mathrm{SNR}=7$ for the cross-correlation method using the triangle constellation and $CR=5$ for the BSD method using the single-L constellation. 
    }\label{fig:dilaton_reach}
\end{figure*}

Assuming a simple linear coupling of the scalar field $\phi$ to the Standard Model, the Lagrangian can be written as~\cite{Hees_PRD2018}:
\begin{equation}\label{eqn:dilaton_lagrangian}
\mathcal{L} \supset 
\frac{\phi}{\Lambda_\gamma} \frac{F_{\mu\nu} F^{\mu\nu}}{4}
- \frac{\phi}{\Lambda_e} m_e \bar{\psi}_e \psi_e,
\end{equation}
where $F_{\mu\nu}$ is the electromagnetic field tensor, $\psi_e$ and $\bar{\psi}_e$ are the electron field and its Dirac conjugate, $m_e$ is the electron rest mass, and $\Lambda_\gamma$ and $\Lambda_e$ denote the scalar DM coupling constants to photons and electrons, respectively.

When $\Lambda_i^{-1}$ is nonzero, the DM field induces modulations in $m_e$ and the fine-structure constant through its coupling to $F_{\mu\nu}$. These variations, in turn, lead to oscillations in both the size and the refractive index of solid materials. More recently, several detection channels---in particular, the coupling of dilatons to the beam splitter---have been proposed to probe these oscillations using GW interferometers~\cite{Grote_2019, Stadnik_PRL2015, Stadnik_2015, Morisaki_PRD2019, Morisaki_PRD2021}.

In GW interferometers, the size of the mirrors and beam splitter will oscillate asymmetrically, inducing a strain $h(f_0)$ given by~\cite{Gottel_PRL2024}:
\begin{equation}\label{eqn:strain}
h(f_0) \simeq 
\left( 
\frac{1}{\Lambda_\gamma} + \frac{1}{\Lambda_e} 
\right)
\frac{\,\sqrt{2\rho_{\mathrm{DM}}}}{2\pi f_0}
\frac{1}{A_{\mathrm{cal}}(f_0)}.
\end{equation}
The quantity $A_{\mathrm{cal}}(f_0)$ is a transfer function relating the DM-induced fluctuations of the optics to the coupling constant $\Lambda^{-1}_{\mathrm{eff}, \mathrm{min}} = \left(\frac{1}{\Lambda_{\gamma}} + \frac{1}{\Lambda_{e}}\right)_{\mathrm{min}}$. It accounts for the effect of finite light travel time~\cite{Morisaki_PRD2021} and encodes the interferometer response to scalar DM~\cite{Gottel_PRL2024}.

\subsection{Dark Photon Dark Matter}

Dark photons are photon-like spin-1 particles that, similar to dilatons, could explain the relic abundance of DM~\cite{AGRAWAL_PRB2020}. These particles could arise from the misalignment mechanism~\cite{Nelson_PRD2011, Arias_2012, Graham_PRD2016}, parametric resonance or the tachyonic instability of a scalar field~\cite{Co_2019, Bastero_Gil_2019, Dror_2019}, or cosmic string network decays~\cite{Long_PRD2019}. The observable effect would result from a coupling of dark photons to Standard Model particles---either (i) to baryons, via $U(1)_B$ gauge symmetry; or (ii) to baryon minus lepton number, via $U(1)_{B-L}$ gauge symmetry. In particular, this interaction would cause ``dark'' forces on the mirrors, inducing oscillations at a frequency fixed by the mass of the dark photon.

\begin{figure*}[hbt!]
    \centering
    \includegraphics[clip,width=\textwidth]{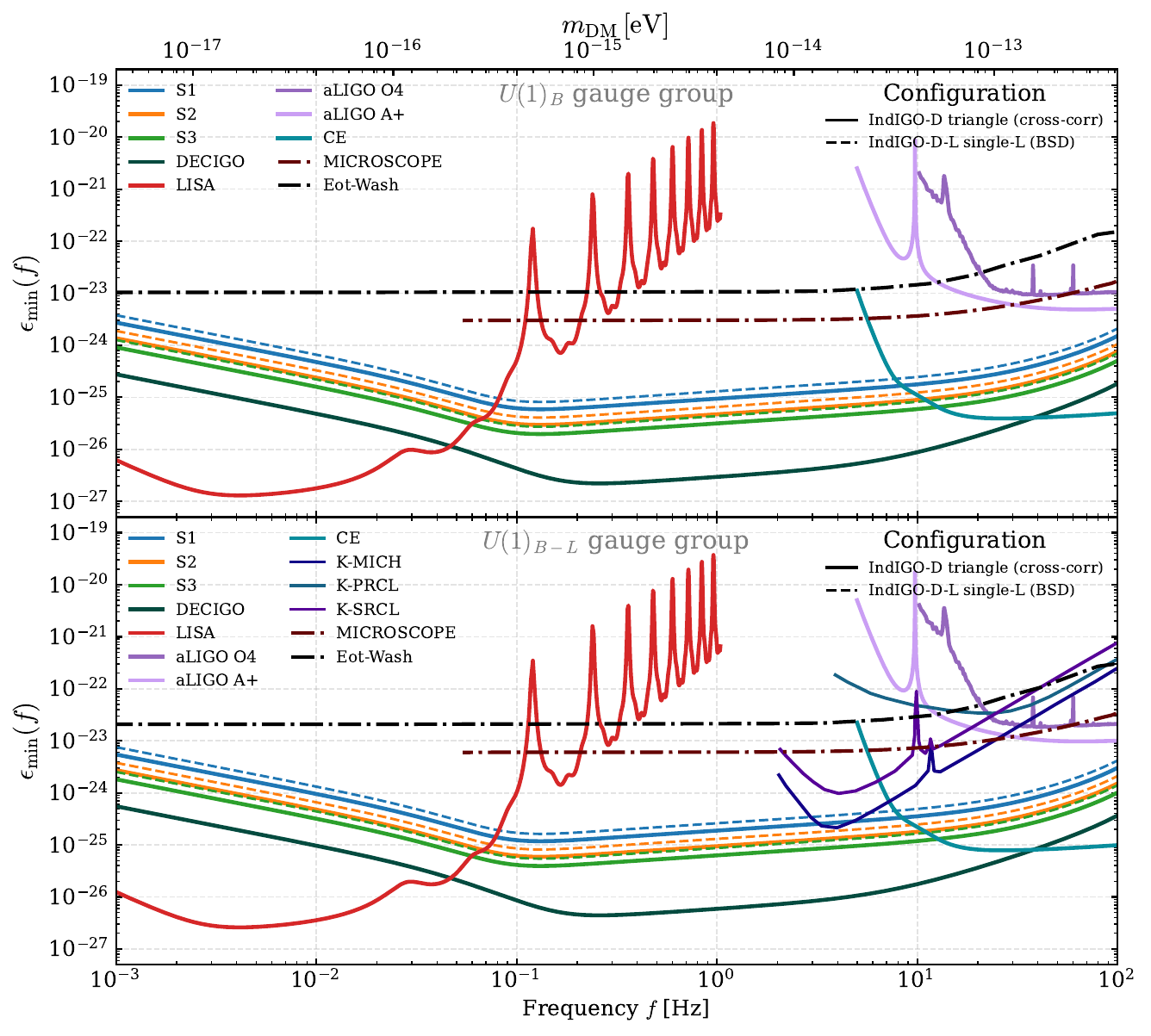}
    \caption{\textit{Top panel:} Projected sensitivity to the dark-photon DM coupling constant $\epsilon_{\mathrm{min}}(f)$ for the $U(1)_B$ gauge group as a function of the dark-photon mass $m_{\mathrm{DM}}$, assuming a continuous observation time of $2$~years. Curves are shown for three representative deciHz GW interferometer noise scenarios (S1, S2, and S3), each under two configurations: (i) a triangular constellation (solid lines) and (ii) a single L-shaped constellation (dashed lines), alongside projections for Advanced-LIGO O4, Advanced-LIGO A+, DECIGO, and LISA. Existing experimental constraints from the MICROSCOPE~\cite{MICROSCOPE_2022} and Eöt-Wash~\cite{EotWash_2008} experiments are also displayed for comparison. Sensitivities are computed assuming a detection threshold of $\mathrm{SNR}=7$ for the cross-correlation method (triangular constellation) and $\mathrm{CR}=5$ for the BSD method (single L-shaped constellation). \textit{Bottom panel:} Projected sensitivity to $\epsilon_{\mathrm{min}}(f)$ for the $U(1)_{B-L}$ gauge group for the same setups. In this case, sensitivity projections derived from KAGRA auxiliary channels (K-PRCL, K-MICH, and K-SRCL)~\cite{ULDM_Kagra_aux} are additionally shown in the high-frequency regime for comparison, alongside all other sensitivity curves evaluated for $U(1)_{B}$.
    }\label{fig:vector_UB_reach}
\end{figure*}

The Lagrangian $\mathcal{L}$ that characterizes the dark photon coupling to a number current density $J_{\mu}$ of baryons ($B$) or baryons minus leptons ($B-L$) is~\cite{AGRAWAL_PRB2020, Pierce_2018}:
\begin{equation}\label{eqn:DPDM_lagrangian}
    \mathcal{L} = -\frac{1}{4}F_{\mu\nu}F^{\mu\nu} + \frac{1}{2} m_{\mathrm{DM}}^2 A_{\mu} A^{\mu} - \epsilon e J_{\mu}A^{\mu}.
\end{equation}
The local amplitude $A_{\mu,0}$ of the dark gauge field, $A_\mu$, can be determined by equating its energy density to the local dark matter density:
\begin{equation}\label{eqn:local_amplitude_from_DM_density}
    \frac{1}{2} m_{\mathrm{DM}}^2 A_{\mu,0} A_0^{\mu} = \rho_{\mathrm{DM}},
\end{equation}
where we take the fiducial local DM density to be $\rho_{\mathrm{DM}} = 0.4\,\mathrm{GeV/cm^3}$. Within one coherence timescale, the field can therefore be expressed as~\cite{Pierce_2018}
\begin{equation}\label{eqn:osc_field}
    A_\mu(t,\mathbf{x}) \simeq A_{\mu,0}\, \sin(m_{\mathrm{DM}} t - \mathbf{k}\!\cdot\!\mathbf{x}),
\end{equation}
where $\mathbf{k} = m_{\mathrm{DM}} \mathbf{v}_0$ is the wave vector of the dark-photon mode.

This oscillating background field exerts a time-varying force on any object carrying the corresponding dark charge. Since dark photon DM is non-relativistic, its electric component---arising from the time derivative of the field---dominates over the magnetic part. The acceleration acting on the $i$-th test mass, located at $\mathbf{x}_i$, is given by~\cite{Pierce_2018}:
\begin{equation}\label{eq:accn}
\begin{split}
    \mathbf{a}_i(t,\mathbf{x}_i)
    &= \frac{\mathbf{F}_i(t,\mathbf{x}_i)}{M_i}
    \simeq \epsilon\, e\, \frac{q_{D,i}}{M_i}\, \partial_t \mathbf{A}(t,\mathbf{x}_i) \\
    &= \epsilon\, e\, \frac{q_{D,i}}{M_i}\, m_{\mathrm{DM}} \mathbf{A}_0 \cos(m_{\mathrm{DM}} t - \mathbf{k}\!\cdot\!\mathbf{x}_i),
\end{split}
\end{equation}
where $\epsilon$ denotes the ratio of the dark-photon coupling strength to the electromagnetic coupling $e$ (Eq.~\eqref{eqn:DPDM_lagrangian}), and $M_i$ and $q_{D,i}$ are the total mass and total dark charge of the $i$-th object, respectively. For a $U(1)_B$ gauge symmetry, $q_D$ counts the baryon number; for $U(1)_{B-L}$, it reduces to the number of neutrons for standard detector materials. In the case of silica, which is typical for interferometer mirrors, $q_D/M = 1~\mathrm{GeV}^{-1}$ for $U(1)_B$ and $q_D/M = 1/2~\mathrm{GeV}^{-1}$ for $U(1)_{B-L}$. 

A GW interferometer is sensitive to the \emph{differential} displacement between pairs of freely suspended test masses located along orthogonal arms. This displacement arises from the small difference in phase of the dark-photon field evaluated at the two test-mass positions, $\mathbf{x}_1$ and $\mathbf{x}_2$.  

Assuming that all mirrors have the same charge-to-mass ratio, integrating Eq.~\eqref{eq:accn} twice over time, and averaging over random polarizations and propagation directions, we obtain the strain on the interferometer caused by a dark-photon DM signal:
\begin{equation}\label{eqn:h_signal}
\begin{split}
    \sqrt{\langle h_D^2 \rangle}
    &= C_{\mathrm{geom}}\, \epsilon\, \frac{q_D e}{M}\, |\mathbf{A}_0|\, v_0 \\
    &= C_{\mathrm{geom}}\, \epsilon\, \frac{q_D e}{M}\, \frac{\sqrt{2\rho_{\mathrm{DM}}}}{2 \pi f_0}\, v_0,
\end{split}
\end{equation}
where $C_{\mathrm{geom}}$ is a geometric factor determined by the interferometer configuration; for L-shaped detectors, $C_{\mathrm{geom}} = \sqrt{2}/3$, whereas for a LISA-like Michelson configuration with a $60^{\circ}$ opening angle, $C_{\mathrm{geom}} = 1/\sqrt{6}$.

A second strain contribution arises due to the so-called ``finite light-travel time'' effect, in which the mirrors move during the time light takes to reach them from the beam splitter. This strain can actually exceed that given in Eq.~\eqref{eqn:h_signal} and is expressed as~\cite{Pierce_2018, Manita_2023}:
\begin{equation}\label{eqn:hc_signal}
    \sqrt{\langle h_C^2 \rangle}
    = \frac{\sqrt{3}}{2}\sqrt{\langle h_D^2 \rangle}\frac{2 \pi f_0 L}{v_0}.
\end{equation}

\subsection{Tensor Dark Matter}

A novel massive spin-2 particle arising from massive gravity theories can modify gravity and serve as a viable DM candidate~\cite{Dubovsky_PRL2005, Babichev_PRD2016, Hassan_JHEP2012}. Scenarios in which massive gravitons comprise all of DM have been widely explored~\cite{Aoki_2016, Manita_2023, Marzola_2018, Babichev_PRD2016, Jain_2022, Aoki_2019, Babichev_2016, Kolb_2023, Aoki_2018}, along with the associated phenomenology and detection methodologies using ground-based or space-based laser interferometers~\cite{Manita_spin2_2023, Armaleo_2021, Manita_2023, Zhang_2025}, atom interferometers~\cite{Blas_2025}, pulsar timing arrays~\cite{Sun_2022, Wu_2023, Armaleo_2020, Xia_2023, Cai_2024}, black-hole superradiant instabilities~\cite{Dias_2023}, and electromagnetic conversion experiments \cite{Gue:2026wqd}. In this work, we consider spin-2 tensor DM arising from Hassan-Rosen bimetric gravity~\cite{Hassan_JHEP2012}, consisting of one massless and one massive graviton, where the massive eigenstate with five degrees of freedom accounts for the dark matter~\cite{Marzola_2018}.

\begin{figure*}[hbt!]
    \centering
    \includegraphics[clip,width=\textwidth]{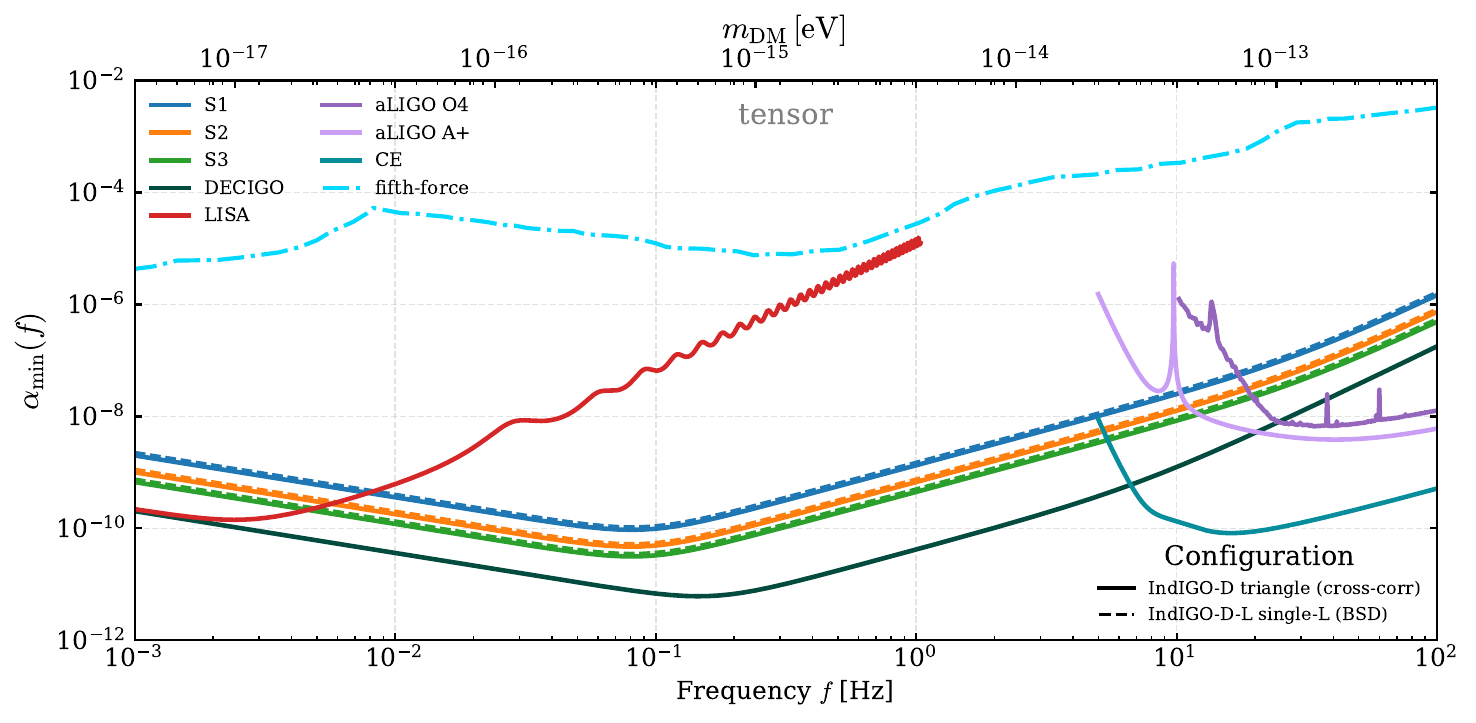} \\ 
    \caption{Projected sensitivity to the DM Yukawa coupling constant $\alpha$ as a function of the tensor boson mass $m_{\rm DM}$, assuming a continuous observation time of 2 years. Curves are shown for three representative proposed deciHz GW interferometers (S1, S2, and S3), each with two different configurations (i) triangular constellation (with solid lines) and (ii) single L-shaped constellation (with dashed lines), as well as for Advanced-LIGO O4, Advanced-LIGO A+, DECIGO, and LISA. Existing constraints from fifth-force experiments~\cite{Armaleo_2021} are also displayed for comparison. The sensitivities are computed for a detection threshold corresponding to $\mathrm{SNR}=7$ for the cross-correlation method using the triangle constellation and $CR=5$ for the BSD method using the single-L constellation.
    }\label{fig:tensor_reach}
\end{figure*}

The Fierz--Pauli Lagrangian density describing the tensor field $\chi_{\mu\nu}$ is given by~\cite{Armaleo_2020}:
\begin{equation}
\mathcal{L}
= -\frac{1}{4}\,\chi_{\mu\nu}\,E^{\mu\nu,\alpha\beta}\,\chi_{\alpha\beta}
 - \frac{1}{8}
 \left( \frac{m_{\mathrm{DM}} c}{\hbar} \right)^{2}
 (\chi_{\mu\nu}\chi^{\mu\nu} - \chi^{2}),
\label{eqn:fierzpauli_lagrangian}
\end{equation}
where $\chi \equiv \chi^{\mu}{}_{\mu}$ and the operator $E^{\mu\nu,\alpha\beta}$ acts on $\chi_{\alpha\beta}$ as
\begin{equation}
\begin{split}
E^{\mu\nu,\alpha\beta}\chi_{\alpha\beta}
= -\tfrac{1}{2}\partial^{2}\chi^{\mu\nu}
 - \tfrac{1}{2}\partial^{\mu}\partial^{\nu}\chi
 + \partial_{\alpha}\partial^{(\nu}\chi^{\mu)\alpha}
 \\ + \tfrac{1}{2}\eta^{\mu\nu}
 \left(\partial^{2}\chi - \partial_{\alpha}\partial_{\beta}\chi^{\alpha\beta}\right),
\label{eqn:Eoperator}
\end{split}
\end{equation}
where parentheses denote symmetrization, $\partial^{2} \equiv \eta^{\rho\sigma}\partial_{\rho}\partial_{\sigma}$, and $\eta^{\mu\nu}$ is the Minkowski metric.

Using the Friedmann--Lemaître--Robertson--Walker (FLRW) background metric, one obtains the late-time equations of motion for this ultralight field. Analogous to GWs, the strain induced on interferometers arises from the oscillatory stretching of spacetime in the presence of the field. In the linear coupling regime, where the massive spin-2 metric mixes weakly with the massless one through a Yukawa coupling $\alpha$, the strain amplitude is~\cite{Aoki_2016, Manita_2023}:

\begin{equation}
\begin{split}
h(f_0)
= \frac{2\alpha\,\Delta\varepsilon}{2\pi f_0 m_{\mathrm{Pl}}}
\sqrt{\frac{\rho_{\mathrm{DM}}}{2}}
\label{eqn:massive_spin2_strain}
\end{split}
\end{equation}
where $\Delta\varepsilon := \varepsilon_{ij}(n^i n^j - m^i m^j)$, with $\mathbf{n}$ and $\mathbf{m}$ denoting the unit vectors along each interferometer arm, $\varepsilon_{ij}$ the polarization tensor of the ultralight field (with five possible polarization states), and $m_{\mathrm{Pl}}$ the reduced Planck mass.

\subsection{Axions}\label{subsubsec:axions}

The axion is a pseudoscalar field originally proposed to solve the strong CP problem in QCD, known as the ``QCD axion''~\cite{PecceiQuinn_PRL1977, Weinberg_PRL1978, Wilczek_PRL1978}. Over the past decades, it has been found that high-energy physics theories, such as string theory, predict a number of axion-like particles arising from the compactification of extra dimensions~\cite{Svrcek_JHEP2006, Arvanitaki_etal_PRD2010, Visinelli_PRD2019}; hereafter we refer to them collectively as ``axions''. With a typical mass $m_a \ll \mathrm{eV}$, an axion behaves as a non-relativistic fluid in the present Universe owing to its oscillatory evolution~\cite{Khlopov_etal_mnras1985}, making it a cosmologically well-motivated cold ULDM candidate. A small but finite coupling between the axion and the photon is a generic prediction of high-energy physics, and provides the observational handle that is exploited here.

The conventional route to probing axions is to leverage axion--photon conversion in a background magnetic field~\cite{Sikivie_PRL1983}, which is the basis of helioscope, haloscope, and light-shining-through-wall experiments~\cite{CAST_NatPhys2017, ADMX_PRL1998, ALPS_NIMA2009, ABRACADABRA_PRL2016}, none of which have reported a detection (for reviews, see~\cite{Irastorza_PPNP2018, Graham_ARNPS2015}). More recently, a class of proposals has emerged that avoids using strong magnetic fields and instead uses optical cavities~\cite{Melissinos_PRL2009, DeRocco_PRD2018, Obata_PRL2018, Liu_PRD2019, Nagano_PRL2019}. These studies exploit the fact that the axion--photon Chern--Simons coupling differentiates the phase velocities of the two circular polarization states of light~\cite{Carroll_etal_PRD1990, Carroll_PRL1998, Andrianov_PLB2010}.

Concretely, the interaction Lagrangian is written as:
\begin{equation}
\begin{split}
\mathcal{L} \supset
\frac{g_{a\gamma}}{4}\, a(t)\, F_{\mu\nu}\tilde{F}^{\mu\nu}
&= g_{a\gamma}\,\dot{a}(t)\, \epsilon^{ijk} A_i \partial_j A_k \\
&\quad + (\text{total derivative}),
\label{eqn:axion_lagrangian}
\end{split}
\end{equation}
where $g_{a\gamma}$ is the axion--photon coupling constant, $a(t)$ is the axion field, $A_\mu$ is the vector potential, $F_{\mu\nu} \equiv \partial_\mu A_\nu - \partial_\nu A_\mu$, and $\tilde{F}^{\mu\nu} \equiv \epsilon^{\mu\nu\rho\sigma} F_{\rho\sigma}/2$ is its Hodge dual. In the temporal ($A_0 = 0$) and Coulomb ($\partial_i A_i = 0$) gauges, decomposing $A_i$ into left- and right-circular modes yields the modified dispersion relation $\omega^2_{L/R} = k^2 (1 \mp g_{a\gamma}\dot{a}/k)$, and hence distinct phase velocities~\cite{Nagano_PRL2019}:
\begin{equation}
c^2_{L/R} = 1 \mp \frac{g_{a\gamma}\dot{a}}{k},
\label{eqn:axion_phasevel}
\end{equation}
where the minus (plus) sign corresponds to $L$ ($R$). The spatial momentum of the axion field is neglected here since the dark matter is non-relativistic. Writing the present-day field as $a(t) = a_0 \cos[m_a t + \delta_\tau(t)]$, with $\delta_\tau$ constant over the coherence time $\tau = 2\pi/(m_a v_0^2)$, Eq.~\eqref{eqn:axion_phasevel} becomes:
\begin{align}
c_{L/R}(t) &\simeq 1 \pm \delta c_0 \sin[m_a t + \delta_\tau(t)], \\
\delta c_0 &= \frac{g_{a\gamma} a_0 m_a}{2k},
\label{eqn:axion_deltac}
\end{align}
with $a_0$ fixed by the local density through $\rho_{\mathrm{DM}} = a_0^2 m_a^2/2$. Following~\citet{Nagano_PRL2019}, for a laser of wavelength $\lambda = 2\pi/k$, this evaluates to
\begin{equation}
\delta c_0 \simeq 1.3\times10^{-24}
\left( \frac{\lambda}{1550~\mathrm{nm}} \right)
\left( \frac{g_{a\gamma}}{10^{-12}~\mathrm{GeV}^{-1}} \right).
\label{eqn:axion_deltac0_num}
\end{equation}

\begin{figure*}[hbt!]
    \centering
    \includegraphics[clip,width=\textwidth]{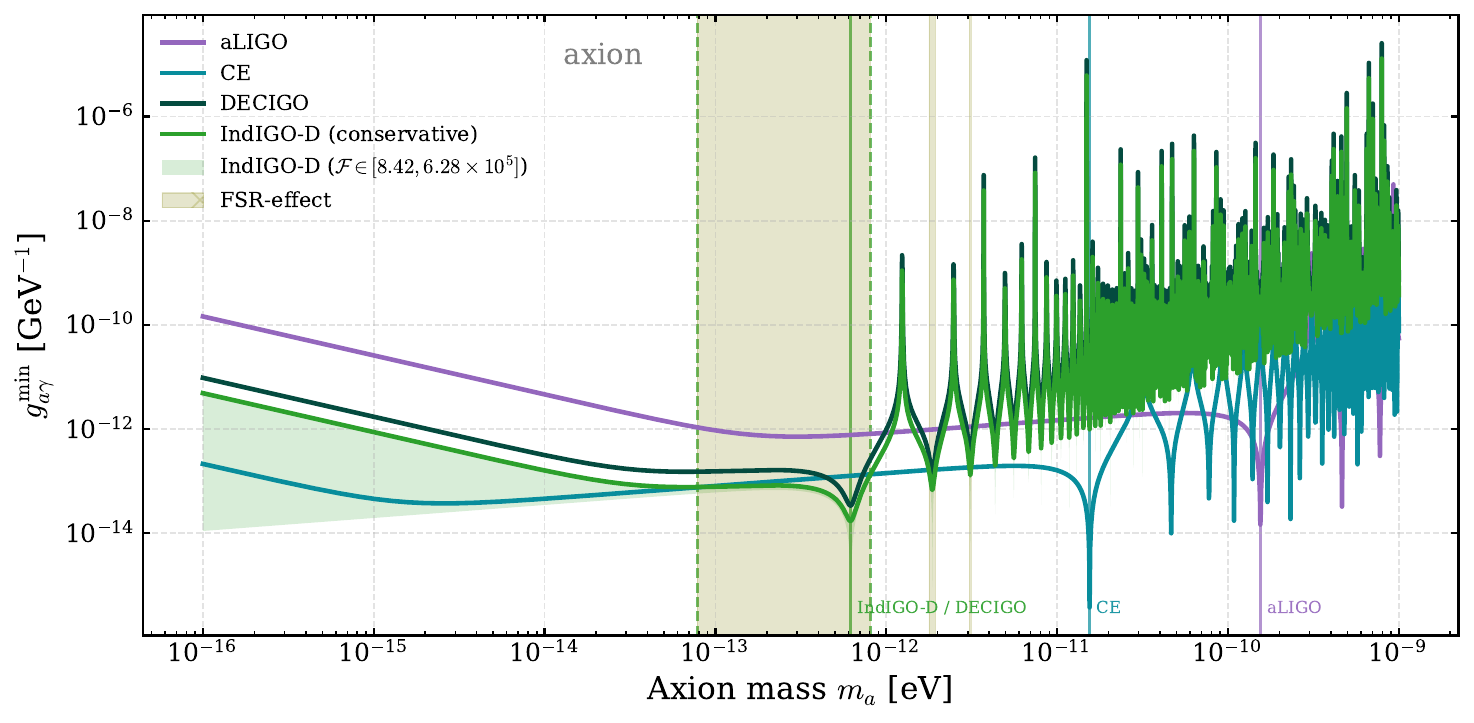} \\ 
    \caption{Projected sensitivity to the axion--photon coupling \(g_{a\gamma}^{\mathrm{min}}\) as a function of the axion mass \(m_a\), assuming a continuous observation time of \(2\) years and shot-noise-limited detector performance. Alongside \indD, curves are shown for Advanced-LIGO, Cosmic Explorer (CE), and DECIGO. Sensitivity peaks at $m=\pi/L$, the free-spectral-range (FSR) resonance mass (marked by dotted horizontal lines) with enhancement proportional to $r_1r_2/(1-r_1r_2)$, are indicated by vertical lines. For \indD (conservative case), we also show the ranges near the FSR-effect, along with its higher harmonics at 3 and 5 times the masses of the fundamental FSR value (see olive-colored bands), where it has an enhanced sensitivity to axion-photon coupling \(g_{a\gamma}^{\mathrm{min}}\). This makes it more sensitive than CE in those specific ranges of axion mass ($m_{a}$). The band of sensitivity reach is illustrated here corresponds to the finesse bound shown in the figure and discussed in Sec.~\ref{subsubsec:axions}. The sensitivity of \indD to axions is only fixed by the specific parameters of the instrument in Eq.~\eqref{eqn:axion_reach}, not by the two pipelines considered for the other types of DM.
    }\label{fig:axion_reach}
\end{figure*}

The observational consequence is that linearly polarized light---a superposition of the two circular states---acquires a small orthogonally polarized component oscillating at the axion frequency. In a linear Fabry--P\'erot cavity of length $L$, resonant at $2kL = 2\pi l$, where $l$ is any positive integer ($l \in \mathbb{N}$), this polarization modulation is resonantly enhanced. The cavity response $H_a(m_a)$, which relates the axion-induced phase-velocity modulation to the amplitude of the orthogonally polarized field at the cavity output, is given by:
\begin{equation}
H_a(m_a) \equiv \frac{i}{k}\, \frac{1}{m_a}\,
\frac{4 r_1 r_2 \sin^2(m_a L/2)}{1 - r_1 r_2 e^{-i 2 m_a L}}
\left( -e^{-i m_a L} \right),
\label{eqn:axion_cavity_response}
\end{equation}
where $r_1$ and $r_2$ are the amplitude reflectivities of the input and end mirrors~\cite{Nagano_PRL2019}. Equation~\eqref{eqn:axion_cavity_response} exhibits three features that shape the sensitivity of interferometers to axions. At $m_a = \pi/L$---the free spectral range---the response is enhanced in proportion to $r_1 r_2/(1 - r_1 r_2)$, producing a sharp sensitivity peak; the same occurs at $m_a L = \pi(2N-1)$, $N \in \mathbb{N}$, though there, $H_a \propto 1/m_a$ because the axion-induced phase accumulated over successive round trips largely cancels, leaving only the contribution from the final half-oscillation during the photon storage time of the cavity. In the opposite, low-mass limit ($m_a L \ll 1$), one finds $H_a \propto m_a$: the effect cancels on the outgoing and returning passes, and the sensitivity degrades. The attainable finesse $\mathcal{F}=\pi\sqrt{r_1 r_2}/(1-r_1 r_2)$, which characterizes the effective number of photon round trips in the cavity, is bounded above by $\sim 10^6$ due to the velocity dispersion of the DM halo~\cite{Millar_JCAP2017}.

A distinctive advantage of this observable is that it is immune to displacement noise. Mirror motion---and indeed a passing GW---imprints a \emph{common} phase shift on the two circular polarizations propagating along the same optical path, which cancels identically in the differential measurement~\cite{Nagano_PRL2019}. The axion search is therefore, in principle, limited only by quantum shot noise, and can proceed without any loss of GW detection sensitivity. The leading source of background noise is mirror roll motion (i.e., rotation of the mirror about its optical axis), which couples to the polarization difference through coating birefringence~\cite{Winkler_OptCommun1994, Aston_CQG2012}.

Assuming only shot noise, the one-sided noise spectrum equivalent to $\delta c$ is
\begin{equation}
\sqrt{S_{\mathrm{shot}}(m_a)} = \frac{\sqrt{k/2P_0}}{\sqrt{\mathcal{T}_j}\, |H_a(m_a)|},
\qquad
\sqrt{\mathcal{T}_j} \equiv \frac{t_1 t_j}{1 - r_1 r_2},
\label{eqn:axion_shot}
\end{equation}
where $P_0$ is the incident power and $j = 2$ ($1$) denotes the transmission (reflection) readout port. The accumulation of SNR depends on whether the observation time exceeds the coherence time~\cite{Budker_PRX2014}, yielding the projected sensitivity reach~\cite{Nagano_PRL2019}:
\begin{equation}
\begin{split}
g_{a\gamma}(m_a) \simeq 1.5\times10^{12}~\mathrm{GeV}^{-1}
\left( \frac{1550~\mathrm{nm}}{\lambda} \right) \\
\times
\begin{cases}
\sqrt{\dfrac{S_{\mathrm{shot}}(m_a)}{T_{\mathrm{obs}}}}, & T_{\mathrm{obs}} \lesssim \tau,\\[10pt]
\sqrt{\dfrac{S_{\mathrm{shot}}(m_a)}{\sqrt{T_{\mathrm{obs}}\tau}}}, & T_{\mathrm{obs}} \gtrsim \tau.
\end{cases}
\label{eqn:axion_reach}
\end{split}
\end{equation}

Note that the $T_{\mathrm{obs}}^{-1/4}$ scaling in the incoherent regime matches that of Eqs.~\eqref{eqn:crosscorr_min_strain_amplitude} and~\eqref{eqn:bsd_min_strain_amplitude}. In every other respect, the axion channel is independent of the strain readout used for the other three ULDM classes: it is a polarimetric rather than a displacement measurement, so the noise model of Eq.~\eqref{eqn:psd} and the scenarios of Table~\ref{tab:psd} do not enter the picture; rather, the sensitivity reach is set directly by the cavity parameters $(L, P_0, \lambda, t_1^2, t_2^2)$ through Eq.~\eqref{eqn:axion_cavity_response}. For the same reason, the mass range in Fig.~\ref{fig:axion_reach} is not restricted to the dHz band of Figs.~\ref{fig:dilaton_reach}--\ref{fig:tensor_reach}: the cavity response is defined at all masses, with a resonant structure at $m_a L = \pi(2N-1)$, $N \in \mathbb{N}$. The signal is enhanced in proportion to $r_1r_2/(1-r_1r_2)$ at $m_a=\pi/L$.

\section{Detection Pipelines: Two Different Search Methods}

We study the effectiveness of two methods to identify different kinds of ULDM that could interact with \indD in the future. These methods are \emph{semi-coherent}, they break the data into segments with durations much shorter than the total observing run and combine the power in each segment without phase information. It is necessary to employ such methods instead of the most sensitive method, matched filtering, because the latter requires a deterministic waveform to correlate with GW data. However, as discussed in Sec.~\ref{sec:uldm}, the signal is coherent only over a timescale $T_{\mathrm{coh}}$, which is significantly shorter than the observation time. Thus, the strategies described below were developed to probe such signals in the context of ground-based GW interferometers, but they can easily be generalized to space-based detectors as well~\cite{Miller:2023kkd}.

\subsection{Cross-Correlation}

With cross-correlation, at least two independent time-series datasets are required. The SNR in this case is evaluated in each frequency bin by dividing the cross-power---obtained by Fourier transforming the time series and multiplying them together---by the standard deviation of the noise.

The cross-power in the $j$th bin for the detector pair $IJ$ is given by~\cite{Pierce_2018, Guo_2019}:
\begin{equation}
S_{IJ,j}=\frac{1}{N_{\mathrm{FFT}}}\sum_{i=1}^{N_{\mathrm{FFT}}}\frac{z_{I,ij}z^*_{J,ij}}{P_{I,ij}P_{J,ij}}
\label{eqn:crosscorr_cross-power}
\end{equation}
where $z_{I,ij}$ and $z_{J,ij}$ are the Fourier transforms of the time-domain data from detectors $I$ and $J$, respectively, $z^*_{J,ij}$ denotes the complex conjugate of $z_{J,ij}$, and $N_{\mathrm{FFT}} = T_{\mathrm{obs}}/T_{\mathrm{FFT}}$ is the total number of FFT segments taken over an observation time $T_{\mathrm{obs}}$ with segment length $T_{\mathrm{FFT}}$. The quantities $P_{I,ij}$ and $P_{J,ij}$ are the power spectral densities of detectors $I$ and $J$.

The variance of the noise data, in the absence of any signal in the $j$th frequency bin, is:
\begin{equation}
\sigma_{IJ, j}^2 = \frac{1}{N_{\mathrm{FFT}}} \left\langle \frac{1}{2P_{I,ij}P_{J,ij}} \right\rangle_{N_{\mathrm{FFT}}} 
\label{eqn:crosscorr_noise_variance}
\end{equation}

The signal-to-noise ratio (SNR) in the $j$th bin is given by:
\begin{equation}
\mathrm{SNR}_{IJ,j} = \frac{S_{IJ,j}}{\sigma_{IJ, j}}.
\label{eqn:crosscorr_SNR}
\end{equation}
Given the SNR, the corresponding minimum detectable strain amplitude can be estimated as:
\begin{equation}
h_{0,j} = \left( \frac{2\mathrm{SNR}}{|\gamma|} \right)^{1/2} \left( \frac{P_{I,j}P_{J,j}}{T_{\mathrm{obs}} T_{\mathrm{FFT}}} \right)^{1/4},
\label{eqn:crosscorr_min_strain_amplitude}
\end{equation}
where $\gamma$ is the overlap reduction function (ORF). The ORF characterizes the sensitivity loss when cross-correlating non-co-located, non-aligned detector pairs.

For the H1 (Hanford) and L1 (Livingston) detector pair, the normalized overlap reduction functions yield $\gamma^{(C)}_{\mathrm{H1L1}} = -0.0297$ and $\gamma^{(D)}_{\mathrm{H1L1}} = -0.89$ for the common ($C$) and differential ($D$) finite light-travel-time strains of vector DM, respectively, whereas $\gamma^{(C)}_{AE} = \gamma^{(D)}_{AE} \simeq -0.756$ for a LISA-like geometry (see Section~\ref{app:orf}). The detector response to tensor DM is similar to the differential vector DM response. For dilatons, only the response arising from the oscillations of the beam splitter and mirrors is considered; hence, it does not involve a geometric ORF.

\subsection{BSD excess power method}

The BSD excess power method uses Band-Sampled Data (BSD) structures~\cite{Piccinni_2019} that allow $T_{\mathrm{FFT}}$ to be varied arbitrarily as a function of frequency. Unlike cross-correlation, it does not require two independent time-series datasets; instead, it relies on generating time-frequency spectrograms for each detector separately, retaining local maxima in frequency bins for each FFT only if their equalized power exceeds a threshold $\theta_{\mathrm{thr}} = 2.5$.

The equalized power in each time and frequency bin is given by the ratio $R_{ij}$ of the squared magnitude of the FFT to a running-median estimation of the power spectral density:
\begin{equation}
R_{ij} = \frac{|\mathrm{FFT}|^2_{I,ij}}{P_{I,ij}},
\label{eqn:bsd_equalized_power}
\end{equation}
which on average is of order $\mathcal{O}(1)$. After applying these cuts, a time-frequency ``peakmap'' is created~\cite{Astone_2005, Astone_prd2014}, consisting of binary values (ones) indicating the specific time-frequency points where the criteria are met. A peakmap can be created at every frequency bin, and $T_{\mathrm{FFT}}$ can be adjusted to match $T_{\mathrm{coh}}$ under the expectation that the signal is nearly sinusoidal. Summing the ones present in each frequency bin constructs a histogram. This procedure is robust against noise disturbances because raw power is not directly summed~\cite{Astone_2005}. From this histogram, at each frequency, we compute a detection statistic called the critical ratio (CR):
\begin{equation}
\mathrm{CR} = \frac{n-\mu}{\sigma},
\label{eqn:bsd_CR}
\end{equation}
where $n$ is the number of triggers beyond the threshold $\theta_{\mathrm{thr}}$ at a given frequency, while $\mu$ and $\sigma$ are the mean and standard deviation of the number of triggers in the histogram, respectively. Like the SNR, the CR follows a normal distribution. Using the CR and assuming Gaussian noise, one can derive the minimum strain amplitude of a sinusoidal signal that, in a frequentist interpretation, produces a detectable signal in a fraction $\ge \Gamma$ of repeated experiments~\cite{Astone_prd2014, Miller_prd2021}:
\begin{equation}
\begin{split}
h_{0,\mathrm{min}} &\approx \frac{\mathcal{G}}{T_{\mathrm{obs}}^{1/4}T_{\mathrm{FFT}}^{1/4}} \sqrt{\frac{P_I(f)}{2}} \\
&\quad \times \left( \frac{p_0(1-p_0)}{p_1^2} \right)^{1/4} \sqrt{\mathrm{CR}_{\mathrm{thr}} - \sqrt{2}\,\mathrm{erfc}^{-1}(2\Gamma)}, \\
p_0 &= e^{-\theta_{\mathrm{thr}}} - e^{-2\theta_{\mathrm{thr}}} + \frac{1}{3}e^{-3\theta_{\mathrm{thr}}}, \\
p_1 &= \frac{\theta_{\mathrm{thr}}}{2}p_0 + \frac{1}{4}e^{-2\theta_{\mathrm{thr}}} - \frac{1}{9}e^{-3\theta_{\mathrm{thr}}},
\label{eqn:bsd_min_strain_amplitude}
\end{split}
\end{equation}
where $\mathcal{G}=\sqrt{\frac{2\pi}{2.4308}}$ arises from convolving a sinusoid with a rectangular window function when computing the sensitivity of GW interferometers to a sinusoidal signal~\cite{Astone_prd2014}.

\section{Geometric factors for the differential vector and tensor responses}\label{geom_factors}

The configuration factor $C_{\mathrm{geom}}$ appearing in Eq.~\eqref{eqn:h_signal} and the root-mean-square (rms) value of $\Delta\varepsilon$ entering Eq.~\eqref{eqn:massive_spin2_strain} are purely geometric quantities: they follow from averaging the detector response over the isotropic random orientations of the DM field. This appendix derives both quantities for a Michelson interferometer with an arbitrary opening angle $\zeta$ between its arm unit vectors, defined by $\cos\zeta \equiv \hat{\mathbf{m}}\cdot\hat{\mathbf{n}}$ ($\zeta=\pi/2$ for an L-shaped geometry and $\zeta=\pi/3$ for a LISA-like geometry).

\paragraph{Isotropic averages.}
Only two averaging identities are required. For a unit vector $\hat{\mathbf{v}}$ distributed uniformly over the sphere, $\langle \hat{v}_a \hat{v}_b\rangle = \frac{1}{3}\delta_{ab}$, so that for any two fixed vectors $\mathbf{p}$ and $\mathbf{q}$,
\begin{equation}
  \big\langle
    (\hat{\mathbf{v}}\cdot\mathbf{p})\,
    (\hat{\mathbf{v}}\cdot\mathbf{q})
  \big\rangle
  = \frac{1}{3}\,\mathbf{p}\cdot\mathbf{q}.
  \label{eq:isoavg_vec}
\end{equation}

For the spin-2 field, the polarization tensor is written in terms of the five polarization matrices, $\varepsilon_{ij} = \sum_{\kappa}\varepsilon_{\kappa}Y^{\kappa}_{ij}$. The $Y^{\kappa}_{ij}$ matrices are orthonormal, $Y^{\kappa}_{ij}Y^{\kappa'}_{ij} = \delta^{\kappa\kappa'}$, and the amplitudes obey $\sum_{\kappa}\varepsilon_{\kappa}^2 = 1$. With no preferred polarization, each of the five modes contributes equally:
\begin{equation}
  \langle \varepsilon_{\kappa}\,\varepsilon_{\kappa'} \rangle
  = \frac{1}{5}\,\delta_{\kappa\kappa'}.
  \label{eq:isoavg_ten}
\end{equation}

\subsection{Differential vector response}\label{app:geom:vec}

Following the derivations in~\citet{Pierce_2018,Manita_2023}, the differential (spatial) pattern function of Eq.~\eqref{eq:FD} is
\begin{equation}
  F^{\lambda}_{D}
  = (\hat{\mathbf{e}}^{\lambda}\!\cdot\hat{\mathbf{m}})
    (\hat{\boldsymbol{\Omega}}\cdot\hat{\mathbf{m}})
  - (\hat{\mathbf{e}}^{\lambda}\!\cdot\hat{\mathbf{n}})
    (\hat{\boldsymbol{\Omega}}\cdot\hat{\mathbf{n}})
  \equiv P - Q.
  \label{eq:FD_PQ}
\end{equation}

For non-relativistic DM, the polarization $\hat{\mathbf{e}}^{\lambda}$ and the propagation direction $\hat{\boldsymbol{\Omega}}$ are \emph{independent} and both isotropic, so every average factorizes into an $\hat{\mathbf{e}}$ factor and an $\hat{\boldsymbol{\Omega}}$ factor, each evaluated using Eq.~\eqref{eq:isoavg_vec}:
\begin{align}
  \langle P^2\rangle
    &= \big\langle(\hat{\mathbf{e}}^{\lambda}\!\cdot\hat{\mathbf{m}})^2\big\rangle
       \big\langle(\hat{\boldsymbol{\Omega}}\cdot\hat{\mathbf{m}})^2\big\rangle
     = \tfrac{1}{3}\cdot\tfrac{1}{3}=\tfrac{1}{9},
  \label{eq:Psq}\\
  \langle Q^2\rangle &= \tfrac{1}{9},
  \label{eq:Qsq}\\
  \langle PQ\rangle
    &= \big\langle(\hat{\mathbf{e}}^{\lambda}\!\cdot\hat{\mathbf{m}})
        (\hat{\mathbf{e}}^{\lambda}\!\cdot\hat{\mathbf{n}})\big\rangle
       \big\langle(\hat{\boldsymbol{\Omega}}\cdot\hat{\mathbf{m}})
        (\hat{\boldsymbol{\Omega}}\cdot\hat{\mathbf{n}})\big\rangle
  \nonumber\\
    &= \Big(\tfrac{\cos\zeta}{3}\Big)\Big(\tfrac{\cos\zeta}{3}\Big)
     = \tfrac{1}{9}\cos^2\zeta.
  \label{eq:PQ}
\end{align}

Assembling these terms into $\langle F_D^2\rangle = \langle P^2\rangle - 2\langle PQ\rangle + \langle Q^2\rangle$ yields:
\begin{equation}
  \langle F_D^2\rangle
  = \tfrac{2}{9}\big(1-\cos^2\zeta\big)
  = \tfrac{2}{9}\sin^2\zeta,
  \label{eq:FDsq}
\end{equation}
so the geometric factor of Eq.~\eqref{eqn:h_signal} is
\begin{equation}
  \boxed{\;
  C_{\mathrm{geom}}(\zeta)
  = \sqrt{\langle F_D^2\rangle}
  = \frac{\sqrt{2}}{3}\,\big|\sin\zeta\big|,
  \;}
  \label{eq:Cgeom}
\end{equation}
which yields $C_{\mathrm{geom}}(\pi/2)=\sqrt{2}/3$ for an L-shaped configuration and $C_{\mathrm{geom}}(\pi/3)=1/\sqrt{6}$ for a $60^{\circ}$ configuration~\cite{Pierce_2018}.

\subsection{Tensor response}\label{app:geom:ten}

Following the derivation in~\citet{Armaleo_2021}, the tensor strain of Eq.~\eqref{eqn:massive_spin2_strain} enters through the projection $\Delta\varepsilon = \varepsilon_{ij}T^{ij}$, where
\begin{equation}
  T_{ij} \equiv \hat{n}_i\hat{n}_j - \hat{m}_i\hat{m}_j
  = 2\,D_{ij}
  \label{eq:Tdef}
\end{equation}
is twice the detector tensor of Eq.~\eqref{eq:dettensor}. Expanding $\varepsilon_{ij}$ in the polarization basis and averaging using Eq.~\eqref{eq:isoavg_ten},
\begin{align}
  \langle \Delta\varepsilon^2 \rangle
    &= \sum_{\kappa\kappa'}
       \langle \varepsilon_{\kappa}\varepsilon_{\kappa'} \rangle\,
       (Y^{\kappa}_{ij}T^{ij})(Y^{\kappa'}_{kl}T^{kl})
  \nonumber\\
    &= \tfrac{1}{5} \sum_{\kappa}\big(Y^{\kappa}_{ij}T^{ij}\big)^2
     = \tfrac{1}{5}\, T_{ij}T^{ij},
  \label{eq:Deps_parseval}
\end{align}
where the last step follows from completeness: $T_{ij}$ is symmetric and trace-free ($T_{ii} = \hat{\mathbf{n}}\cdot\hat{\mathbf{n}} - \hat{\mathbf{m}}\cdot\hat{\mathbf{m}} = 0$), so it is spanned by the five orthonormal $Y^{\kappa}$ matrices, and the sum of its squared components returns its norm. The norm is evaluated via dot products:
\begin{align}
  T_{ij}T^{ij}
    &= (\hat{\mathbf{n}}\cdot\hat{\mathbf{n}})^2
     + (\hat{\mathbf{m}}\cdot\hat{\mathbf{m}})^2
     - 2(\hat{\mathbf{m}}\cdot\hat{\mathbf{n}})^2
  \nonumber\\
    &= 2 - 2\cos^2\zeta = 2\sin^2\zeta,
  \label{eq:Tnorm}
\end{align}
yielding:
\begin{equation}
  \langle \Delta\varepsilon^2 \rangle
  = \tfrac{1}{5}\big(2\sin^2\zeta\big)
  = \tfrac{2}{5}\,\sin^2\zeta.
  \label{eq:Depssq}
\end{equation}
Thus, the rms polarization projection is:
\begin{equation}
  \boxed{\;
  \sqrt{\langle \Delta\varepsilon^2\rangle}
  = \sqrt{\frac{2}{5}}\,\big|\sin\zeta\big|.
  \;}
  \label{eq:Deps_rms}
\end{equation}
Numerically, $\sqrt{\langle\Delta\varepsilon^2\rangle} = \sqrt{2/5} \simeq 0.632$ at $\zeta=\pi/2$ and $\sqrt{3/10}\simeq0.548$ at $\zeta=\pi/3$; the rms tensor strain follows from Eq.~\eqref{eqn:massive_spin2_strain} with $\Delta\varepsilon \to \sqrt{2/5}\,|\sin\zeta|$. Here, we present the first derivation of the geometric factor for tensor bosons in a triangular configuration.

\section{Overlap Reduction Functions (ORFs) for dark photons}\label{app:orf}

\subsection{ORF for L-shaped geometry} 

A dark photon field exerts a force proportional to $\dot{\mathbf{A}}$ on a test mass. The induced acceleration points along the instantaneous field direction, which we denote by the unit vector $\hat{\mathbf{e}}^{\lambda}$ (the polarization). Both mirrors of a given arm are displaced in the \emph{same} direction (common motion); two mechanisms then generate a measurable change in arm-length (see~\citet{Manita_2023} for full details of the calculation in this subsection).

\paragraph{Time (finite light-travel) response.}
During the laser round trip, the mirrors move, and the measured phase responds to the mirror displacement projected along the arm. For an arm with unit vector $\hat{\mathbf{n}}$, the single-arm response is proportional to
\begin{equation}
  r^{\mathrm{time}}_{\mathrm{arm}}(\hat{\mathbf{e}}^{\lambda}) \;\propto\; \hat{\mathbf{e}}^{\lambda}\cdot\hat{\mathbf{n}}.
  \label{eq:armresp_t}
\end{equation}
A Michelson interferometer measures the difference between its two arms, defined by unit vectors $\hat{\mathbf{m}}$ and $\hat{\mathbf{n}}$. The polarization vector $\hat{\mathbf{e}}^{\lambda}$ is common to both arms and factors out:
\begin{equation}
\begin{split}
  R^{\mathrm{time}}(\hat{\mathbf{e}}^{\lambda})
    &= \hat{\mathbf{e}}^{\lambda}\cdot(\hat{\mathbf{m}}-\hat{\mathbf{n}})
     = \hat{\mathbf{e}}^{\lambda}\cdot\mathbf{D}, \\
  \mathbf{D} &\equiv \hat{\mathbf{m}} - \hat{\mathbf{n}}.
\end{split}
\label{eq:detvec}
\end{equation}
The geometry enters through the \emph{detector vector} $\mathbf{D}$, defined as the difference between the two arm unit vectors.

\paragraph{Space (differential) response.}
Because the field has a finite velocity $v$, its value differs slightly between the two ends of an arm; the resulting differential force leaves a residual arm-length change. This response carries an additional factor of the field's propagation direction $\hat{\boldsymbol{\Omega}}$ projected onto the arm, so the single-arm response is quadratic in $\hat{\mathbf{n}}$,
\begin{equation}
  r^{\mathrm{space}}_{\mathrm{arm}}(\hat{\mathbf{e}}^{\lambda},\hat{\boldsymbol{\Omega}})
    \;\propto\; (\hat{\mathbf{e}}^{\lambda}\cdot\hat{\mathbf{n}})\,
                (\hat{\boldsymbol{\Omega}}\cdot\hat{\mathbf{n}}).
  \label{eq:armresp_s}
\end{equation}
Taking the difference between the two arms, the detector geometry enters through the \emph{detector tensor}
\begin{equation}
\begin{split}
  R^{\mathrm{space}}(\hat{\mathbf{e}}^{\lambda},\hat{\boldsymbol{\Omega}})
    &= 2\,\hat{e}^{\lambda}_a\, D_{ab}\, \hat{\Omega}_b, \\
  D_{ab} &\equiv \tfrac{1}{2}\left(\hat{m}_a\hat{m}_b - \hat{n}_a\hat{n}_b\right).
\end{split}
  \label{eq:dettensor}
\end{equation}
The two corresponding pattern functions for the common ($C$) and differential ($D$) modes are given by
\begin{align}
  F^{\lambda}_{C}(\hat{\boldsymbol{\Omega}})
    &= (\hat{\mathbf{m}}-\hat{\mathbf{n}})\cdot\hat{\mathbf{e}}^{\lambda},
    \label{eq:FC}\\
  F^{\lambda}_{D}(\hat{\boldsymbol{\Omega}})
    &= (\hat{\mathbf{m}}\otimes\hat{\mathbf{m}}
       - \hat{\mathbf{n}}\otimes\hat{\mathbf{n}})_{ab}\,
       (\hat{\mathbf{e}}^{\lambda}\otimes\hat{\boldsymbol{\Omega}})^{ab}.
    \label{eq:FD}
\end{align}
The ORF between detectors $I$ and $J$ is the normalized, polarization-summed solid-angle average of the pattern functions,
\begin{equation}
\begin{split}
  \gamma^{(s)}_{IJ}
    &= \frac{\Gamma^{(s)}_{IJ}}{\sqrt{\Gamma^{(s)}_{II}\,\Gamma^{(s)}_{JJ}}}, \\
  \Gamma^{(s)}_{IJ}
    &= \sum_{\lambda}\frac{1}{4\pi}\!\int d\hat{\boldsymbol{\Omega}}\,
      F^{\lambda *}_{s,I}\,F^{\lambda}_{s,J},
  \qquad s=C,D.
\end{split}
  \label{eq:orfdef}
\end{equation}

In the long-wavelength limit, the angular integrals reduce to simple contractions of the arm vectors. For the common mode, this is the normalized dot product of the \emph{detector vectors} $\mathbf{D}\equiv\hat{\mathbf{m}}-\hat{\mathbf{n}}$,
\begin{equation}
  \gamma^{(C)}_{IJ}
    = \frac{\mathbf{D}_I\cdot\mathbf{D}_J}
           {|\mathbf{D}_I|\,|\mathbf{D}_J|},
  \label{eq:gC}
\end{equation}
while for the differential mode, it is the normalized contraction of the \emph{detector tensors} $D_{ab}\equiv\tfrac{1}{2}(\hat{m}_a\hat{m}_b-\hat{n}_a\hat{n}_b)$,
\begin{equation}
  \gamma^{(D)}_{IJ}
    = \frac{D_I^{ab}\,D_{J,ab}}
           {\sqrt{D_I^{ab}D_{I,ab}\;D_J^{cd}D_{J,cd}}}.
  \label{eq:gD}
\end{equation}
Using the H1 and L1 arm unit vectors,
\begin{align}
  \hat{\mathbf{m}}_{\mathrm{H1}} &= (-0.2239,\ 0.7998,\ 0.5569), \\
  \hat{\mathbf{n}}_{\mathrm{H1}} &= (-0.9140,\ 0.0261,\ -0.4049), \\
  \hat{\mathbf{m}}_{\mathrm{L1}} &= (-0.9546,\ -0.1416,\ -0.2622), \\
  \hat{\mathbf{n}}_{\mathrm{L1}} &= (0.2977,\ -0.4879,\ -0.8205),
  \label{eq:ligoarms}
\end{align}
Eqs.~\eqref{eq:gC} and \eqref{eq:gD} yield
\begin{equation}
  \gamma^{(C)}_{\mathrm{H1L1}} = -0.0297,
  \qquad
  \gamma^{(D)}_{\mathrm{H1L1}} = -0.89.
  \label{eq:ligovals}
\end{equation}

\subsection{ORF for triangular geometry}

For a LISA-like geometry of detectors in a triangular constellation, the three vertices share laser links and have correlated noise; thus, cross-correlation is performed between noise-orthogonal synthetic channels. For concreteness, we place one vertex at the origin and orient the unit-side triangle in the $xy$ plane~\footnote{The result is independent of this placement: the ORFs depend only on the relative arm directions, which are unchanged by a translation or rotation of the whole constellation.},
\begin{equation}
  \mathbf{P}_Z = (0,0,0),\quad
  \mathbf{P}_X = (1,0,0),\quad
  \mathbf{P}_Y = \big(\tfrac{1}{2},\frac{\sqrt{3}}{2},0\big).
  \label{eq:lisapos}
\end{equation}
Each vertex is a $60^\circ$ Michelson interferometer whose two arms are the unit vectors toward the other two spacecraft. The arm unit vectors are given by
\begin{align}
  \hat{\mathbf{m}}_X &= \big(-\tfrac{1}{2},\ \frac{\sqrt{3}}{2},\ 0\big), &
  \hat{\mathbf{n}}_X &= (-1,\ 0,\ 0), \\
  \hat{\mathbf{m}}_Y &= \big(\tfrac{1}{2},\ -\frac{\sqrt{3}}{2},\ 0\big), &
  \hat{\mathbf{n}}_Y &= \big(-\tfrac{1}{2},\ -\frac{\sqrt{3}}{2},\ 0\big), \\
  \hat{\mathbf{m}}_Z &= (1,\ 0,\ 0), &
  \hat{\mathbf{n}}_Z &= \big(\tfrac{1}{2},\ \frac{\sqrt{3}}{2},\ 0\big),
\end{align}
yielding the detector vectors
\begin{equation}
  \mathbf{D}_X = \big(\tfrac{1}{2},\frac{\sqrt{3}}{2},0\big),\quad
  \mathbf{D}_Y = (1,0,0),\quad
  \mathbf{D}_Z = \big(\tfrac{1}{2},-\frac{\sqrt{3}}{2},0\big),
  \label{eq:lisavecs}
\end{equation}
with the detector tensors $D^{(X,Y,Z)}_{ab}$ formed analogously from the same arms. The noise-orthogonal channels are fixed linear combinations of the three vertices, as in Eq.~(9.12) of~\citet{Romano_2017}:
\begin{equation}
  A = \tfrac{1}{3}\left(2X - Y - Z\right),
  \qquad
  E = \frac{1}{\sqrt{3}}\left(Z - Y\right),
  \label{eq:AEdef}
\end{equation}
and the corresponding detector vectors/tensors are formed using the same combinations (e.g., $\mathbf{D}_A = \tfrac{1}{3}(2\mathbf{D}_X-\mathbf{D}_Y-\mathbf{D}_Z) = \big(-\tfrac{1}{6},\frac{\sqrt{3}}{2},0\big)$ and $\mathbf{D}_E = \frac{1}{\sqrt{3}}(\mathbf{D}_Z-\mathbf{D}_Y) = \big(-\frac{1}{2\sqrt{3}},-\tfrac{1}{2},0\big)$). This yields the common-mode finite light-travel time response ORF for the LISA-like geometry:
\begin{equation}
\boxed{
    \gamma^{(C)}_{AE} \simeq -0.756
}
  \label{eq:aetorf_C}
\end{equation}
Here, we derive for the first time the overlap reduction function (ORF) for the common-mode motion effect in triangular interferometers.

The differential-mode detector tensor is
\begin{equation}
D_{ab}
=\frac{1}{2}\left(
\hat m_a\hat m_b-\hat n_a\hat n_b
\right).
\end{equation}

Using the arm vectors defined above, the detector tensors at the three
vertices are

\begin{align}
D_X
&=\frac{1}{2}\!\left[
\left(-\tfrac{1}{2},\frac{\sqrt{3}}{2},0\right)^{\!\otimes2}
-(-1,0,0)^{\otimes2}
\right] \nonumber\\
&=
\begin{pmatrix}
-\tfrac{3}{8} & -\frac{\sqrt{3}}{8} & 0\\[2pt]
-\frac{\sqrt{3}}{8} & \tfrac{3}{8} & 0\\[2pt]
0 & 0 & 0
\end{pmatrix},
\label{eq:DX}
\\[4pt]
D_Y
&=\frac{1}{2}\!\left[
\left(\tfrac{1}{2},-\frac{\sqrt{3}}{2},0\right)^{\!\otimes2}
-\left(-\tfrac{1}{2},-\frac{\sqrt{3}}{2},0\right)^{\!\otimes2}
\right] \nonumber\\
&=
\begin{pmatrix}
0 & -\frac{\sqrt{3}}{4} & 0\\[2pt]
-\frac{\sqrt{3}}{4} & 0 & 0\\[2pt]
0 & 0 & 0
\end{pmatrix},
\label{eq:DY}
\\[4pt]
D_Z
&=\frac{1}{2}\!\left[
(1,0,0)^{\otimes2}
-\left(\tfrac{1}{2},\frac{\sqrt{3}}{2},0\right)^{\!\otimes2}
\right] \nonumber\\
&=
\begin{pmatrix}
\tfrac{3}{8} & -\frac{\sqrt{3}}{8} & 0\\[2pt]
-\frac{\sqrt{3}}{8} & -\tfrac{3}{8} & 0\\[2pt]
0 & 0 & 0
\end{pmatrix}.
\label{eq:DZ}
\end{align}

The synthetic-channel tensors are the same linear combinations as in
Eq.~\eqref{eq:AEdef}:
\begin{align}
D_A
&=\frac{1}{3}\left(2D_X-D_Y-D_Z\right)\nonumber\\
&=
\begin{pmatrix}
-\tfrac{3}{8} & \frac{\sqrt{3}}{24} & 0\\[2pt]
\frac{\sqrt{3}}{24} & \tfrac{3}{8} & 0\\[2pt]
0 & 0 & 0
\end{pmatrix},
\label{eq:DAtensor}
\\[4pt]
D_E
&=\frac{1}{\sqrt{3}}\left(D_Z-D_Y\right)\nonumber\\
&=
\begin{pmatrix}
\frac{\sqrt{3}}{8} & \tfrac{1}{8} & 0\\[2pt]
\tfrac{1}{8} & -\frac{\sqrt{3}}{8} & 0\\[2pt]
0 & 0 & 0
\end{pmatrix}.
\label{eq:DEtensor}
\end{align}

The required contractions
$D_I:D_J\equiv D_I^{ab}D_{J,ab}$ (the sum of element-wise products of the two matrices) are
\begin{equation}
  D_A\!:\!D_A = \tfrac{7}{24},
  \qquad
  D_E\!:\!D_E = \tfrac{1}{8},
  \qquad
  D_A\!:\!D_E = -\frac{1}{4\sqrt{3}},
  \label{eq:aecontractions}
\end{equation}
so that, from Eq.~\eqref{eq:gD},
\begin{equation}
\begin{split}
  \gamma^{(D)}_{AE}
    = \frac{D_A\!:\!D_E}
           {\sqrt{(D_A\!:\!D_A)(D_E\!:\!D_E)}} \\
    = \frac{-1/(4\sqrt3)}
           {\sqrt{(7/24)(1/8)}}\\
           \boxed{\;
    \gamma^{(D)}_{AE} = -\sqrt{\tfrac{4}{7}}
    \simeq -0.756,
    \;}
  \label{eq:aetorf}
\end{split}
\end{equation}
which is equal to the common-mode value $\gamma^{(C)}_{AE}$ obtained above. Note that our derivation of the differential-mode ORF for triangular interferometers differs from the value of $-0.29$ reported by \citet{Pierce_2018}.

\section{Results}\label{sec:results}

We present projected detection sensitivities for all four widely proposed classes of non-relativistic, bosonic ULDM candidates over a continuous observation period of $T_{\mathrm{obs}} = 2~\text{years}$, computed with both search pipelines: (i) cross-correlation between the two quasi-independent channels of the triangular ($60^{\circ}$) configuration, evaluated at a detection threshold of $\mathrm{SNR} = 7$; and (ii) the BSD excess-power statistic applied to the single L-shaped channel, evaluated at a critical-ratio threshold $\mathrm{CR}_{\mathrm{thr}} = 5$, confidence level $\Gamma = 0.95$, and peakmap threshold $\theta_{\mathrm{thr}} = 2.5$ [Eq.~\eqref{eqn:bsd_min_strain_amplitude}].

We find that the detection sensitivities agree to within an order of magnitude across all the classes: (1) scalar dilatons, (2a) $U(1)_{B}$ vector dark photons, (2b) $U(1)_{B-L}$ vector dark photons, and (3) tensor DM (evaluated across three displacement-coupled channels using two distinct pipelines tailored to their respective constellation shapes). 

\subsection{Dilatons}

For the scalar (dilaton) channel, the effective detection sensitivity reach for $\Lambda^{-1}_{\mathrm{eff,min}} \equiv (1/\Lambda_\gamma + 1/\Lambda_e)_{\mathrm{min}}$ is shown in Fig.~\ref{fig:dilaton_reach}, assuming a dimensionless fiducial value of $A_{\mathrm{cal}} = 10^{6}$ in the absence of a computed calibration response for \indD. The sensitivity curve achieves maximum reach near the low-frequency corner of the band. Using \indD's S3 noise PSD, the projected detection sensitivity reaches $\Lambda^{-1}_{\mathrm{eff,min}} \simeq 3.3 \times 10^{-24}~\mathrm{GeV}^{-1}$ at $f \simeq 0.08~\mathrm{Hz}$ ($m_{\mathrm{DM}} \simeq 3.4 \times 10^{-16}~\mathrm{eV}$). For the S1 and S2 noise PSDs, the detection sensitivity at this frequency is weaker by factors of $3.0$ and $1.5$, respectively.

\subsection{Dark photons}

For vector (dark-photon) DM (Fig.~\ref{fig:vector_UB_reach}), the detection sensitivity to the dark-photon coupling parameter $\epsilon_{\mathrm{min}}$ peaks at $f \simeq 0.13~\mathrm{Hz}$ ($m_{\mathrm{DM}} \simeq 5.4 \times 10^{-16}~\mathrm{eV}$) for both the $U(1)_B$ and $U(1)_{B-L}$ gauge groups. Under $U(1)_B$, the S3 noise PSD attains $\epsilon_{\mathrm{min}} \simeq 2.0 \times 10^{-26}$, whereas S2 and S1 reach $2.9 \times 10^{-26}$ and $5.9 \times 10^{-26}$, respectively. Under the $U(1)_{B-L}$ gauge group, the sensitivity is weaker by a factor of two, as expected from the reduced charge-to-mass ratio ($q_D/M = 0.5~\mathrm{GeV}^{-1}$ versus $1~\mathrm{GeV}^{-1}$) found in standard materials, with detection sensitivity for the S3 noise PSD reaching $\epsilon_{\mathrm{min}} \simeq 3.9 \times 10^{-26}$. DECIGO improves on this by roughly an order of magnitude ($2.2 \times 10^{-27}$ for $U(1)_B$), and LISA reaches comparable couplings ($1.3 \times 10^{-27}$) but at much lower masses.

Across the full decihertz band, the S1 noise PSD already improves on the equivalence-principle bounds from the MICROSCOPE~\cite{MICROSCOPE_2022} and E\"ot-Wash~\cite{EotWash_2008} experiments by up to factors of~$51$ and~$177$, respectively. For the S3 PSD, the corresponding improvements reach as high as~$153$ and~$531$. In the high-frequency overlap for the $U(1)_{B-L}$ gauge group, the \indD (S3 PSD) sensitivity is also stronger than the KAGRA auxiliary-channel projections (K-MICH, K-PRCL, K-SRCL)~\cite{ULDM_Kagra_aux}, which attain a peak sensitivity of $\epsilon_{\mathrm{min}} \simeq 2 \times 10^{-25}$ near $4~\mathrm{Hz}$ (see the bottom panel of Fig.~\ref{fig:vector_UB_reach}).

\subsection{Tensor Bosons}

As demonstrated in Fig.~\ref{fig:tensor_reach}, for tensor (massive spin-2 field) DM, the sensitivity reach on the Yukawa coupling $\alpha_{\mathrm{min}}$ again peaks near $0.08~\mathrm{Hz}$ ($m_{\mathrm{DM}} \simeq 3.4 \times 10^{-16}~\mathrm{eV}$), where the S3 noise PSD achieves $\alpha_{\mathrm{min}} \simeq 3.1 \times 10^{-11}$, while the S2 and S1 PSDs attain $4.7 \times 10^{-11}$ and $9.4 \times 10^{-11}$, respectively. By comparison, the detection sensitivity for DECIGO is roughly a factor of five stronger, reaching $6.1 \times 10^{-12}$. The sensitivity for the S3 noise PSD improves on the fifth-force constraint~\cite{Armaleo_2021} (which sits at $\alpha \sim 10^{-5}$ near $0.1~\mathrm{Hz}$) by up to a factor of $5.4 \times 10^{5}$, opening more than five orders of magnitude of previously unconstrained coupling space in the $10^{-16}$--$10^{-13}~\mathrm{eV}$ mass window.

\subsection{Axions}

For aLIGO, CE, and DECIGO, we adopt the parameters from Table~I of Ref.~\cite{Nagano_PRL2019}. For \indD, we take $L = 1000~\mathrm{km}$, incident power $P_0 = 10~\mathrm{W}$, and wavelength $\lambda = 1~\mu\mathrm{m}$, consistent with the instrumental specifications outlined in Sec.~\ref{sec:detector}. The two mirrors are taken to be identical ($t_1 = t_2$), so that the cavity is critically coupled: the prompt reflection off the input mirror and the intracavity leakage cancel out on resonance, leaving no carrier in the reflection port, and the polarization readout must therefore be taken in transmission as described by Eq.~\eqref{eqn:axion_shot}. This is the configuration adopted for DECIGO~\cite{Nagano_PRL2019} and is the natural one for a long-baseline space mission. Ground-based detectors instead use strongly overcoupled arm cavities ($t_1 \gg t_2$) to build up an effective baseline far exceeding their physical arm length; for an interferometer with $\sim 1000~\mathrm{km}$ arms, no such enhancement is required.

Since the mirror coatings are not yet determined, we bracket the projected sensitivity reach over a range of the cavity finesse $\mathcal{F} = \pi\sqrt{r_1 r_2}/(1-r_1 r_2)$. We use the DECIGO transmissivity $t_1^2 = t_2^2 = 3.1\times10^5~\mathrm{ppm}$ ($\mathcal{F} \approx 8.4$) for the conservative estimate (see Fig.~\ref{fig:axion_reach}), and for the optimistic edge we take $t_1^2 = t_2^2 = 5~\mathrm{ppm}$ ($\mathcal{F} \approx 6.3\times10^5$), the end-mirror transmissivity adopted for aLIGO and CE in Table~I of~\citet{Nagano_PRL2019}. Finesse in any case cannot exceed $\mathcal{F} \sim 10^{6}$, at which the cavity linewidth becomes comparable to the DM velocity dispersion and the resonant buildup dissipates the signal coherence~\cite{Nagano_PRL2019, Millar_JCAP2017}.

The axion channel is a polarimetric, shot-noise-limited measurement, independent of the strain readout and noise PSD scenarios discussed above. Instead, sensitivity to axions is set by the cavity parameters through the transfer function described in Eq.~\eqref{eqn:axion_cavity_response}. Over the finesse range quoted above, the conservative edge attains $g_{a\gamma}^{\mathrm{min}} \simeq 1.7\times10^{-14}~\mathrm{GeV}^{-1}$ at the first free-spectral-range resonance $m_a = \pi/L \simeq 6.2\times10^{-13}~\mathrm{eV}$, improving to $4.8\times10^{-16}~\mathrm{GeV}^{-1}$ at the optimistic edge (see Fig.~\ref{fig:axion_reach}). Well below this resonance, where $m_a L \ll 1$ and $H_a \propto m_a$, the response scales with the baseline, so \indD substantially outperforms aLIGO across the mass range shown in Fig.~\ref{fig:axion_reach}. At $m_a = 10^{-15}~\mathrm{eV}$, the conservative detection sensitivity can reach $8.7\times10^{-13}~\mathrm{GeV}^{-1}$, which is a factor of $\sim 30$ deeper than aLIGO ($2.6\times10^{-11}~\mathrm{GeV}^{-1}$) and within a factor of two of DECIGO.

\section{Summary \& discussion}

This study provides a timely and quantitative benchmark for the direct DM detection capabilities of \indD, a proposed space-based GW observatory operating in the decihertz ($\sim 0.01$--$10~\mathrm{Hz}$) regime~\cite{Sharma_2026}. While the primary motivation for decihertz interferometers often centers on capturing early inspirals of compact binary coalescences~\cite{LISA-TAIJI}, our work demonstrates that \indD inherently acts as a powerful, multi-channel direct detector for cold ULDM candidates~\cite{Miller_2026, Ferreira_2021}. We summarize below the key impact and broader importance of our findings.

\indD can bridge the sensitivity gap between space-based and ground-based interferometers for ULDM that induces time-varying displacements of the test masses. Due to the mass-frequency relation in Eq.~\eqref{eqn:f0_freq}, ground-based detectors (LIGO, Virgo, KAGRA, ET, CE)~\cite{LVK_2015, Virgo_2014, Kagra_2020, ET_2026, CE_2023} and low-frequency space missions (LISA, Taiji)~\cite{LISA_2017, Taiji_2023} leave a distinct mass coverage gap (see Figs.~\ref{fig:dilaton_reach}, \ref{fig:vector_UB_reach}, and \ref{fig:tensor_reach}). We show that \indD effectively closes this window, opening access to previously unconstrained coupling parameter spaces for vector ($U(1)_B$ and $U(1)_{B-L}$ dark photons)~\cite{AGRAWAL_PRB2020, Pierce_2018} and tensor DM candidates across $m_{\mathrm{DM}} \sim 10^{-16}$--$10^{-13}~\mathrm{eV}$~\cite{Armaleo_2021, Manita_2023, Aoki_2016}.

By directly comparing the sensitivities of two competing constellation layouts (a single L-shaped Michelson interferometer and a LISA-like triangular geometry with a $60^\circ$ inter-arm angle)~\cite{Sharma_2026, Romano_2017}, evaluated using two independent analysis techniques (cross-correlation and the BSD excess-power method)~\cite{Pierce_2018, Guo_2019, Miller_prd2021, Astone_prd2014}, we establish that our projected coupling limits agree up to an $\mathcal{O}(1)$ factor. This indicates that the mission's DM discovery potential is robust against future methodological and instrumental design choices.

Beyond the primary strain readout, additional instrumental channels for polarization measurement as prescribed in \citet{Nagano_PRL2019}, if accessible, can provide competitive or even unmatched constraints near the free-spectral-range resonance mass set by the baseline ($m_a \sim \pi/L$). These measurements do not affect the nominal GW measurements and, in turn, are unaffected by any noise induced in the strain channel. Moreover, future space-based detectors can be tuned to search for axion DM with only small modifications~\cite{Gue:2024dcu}, providing competitive bounds below the masses probed by \indD.

For axion DM detection, the sensitivity profile is governed by two distinct mechanisms: (i)~the underlying baseline sensitivity floor, and (ii)~a series of sharp transfer-function resonance peaks (comprising the primary free-spectral-range [FSR] frequency and its higher harmonics, see figure~\ref{fig:axion_reach}), which manifest as pronounced local dips in the sensitivity floor~\cite{Nagano_PRL2019}. While the baseline floor is predominantly dictated by core optical parameters, such as cavity mirror reflectivities, laser power, and wavelength, the precise location of the FSR resonant dips is set directly by the interferometer arm length ($L$). Ultimately, the combined reflectivity of the cavity mirrors modulates the overall cavity finesse, thereby tuning the depth and sharpness of these resonant enhancements across the parameter space.

The sensitivities reported here are based on current baseline instrument specifications. Crucially, the ULDM discovery reach of a decihertz mission is not dictated by its macro-constellation geometry alone; it also hinges on subsystem choices—such as mirror substrate composition, auxiliary optical layout, laser power, and operating wavelength. Optimizing these parameters offers substantial scope for sensitivity enhancements without compromising primary compact binary coalescence (CBC) science objectives. For vector dark photons, this material dependence enters explicitly into our formalism, as the strain signal in Eq.~\eqref{eqn:h_signal} scales linearly with the dark charge-to-mass ratio $q_D/M$ of the test masses. Under a $U(1)_B$ symmetry, this ratio is virtually material-independent ($q_D/M \simeq 1~\mathrm{GeV}^{-1}$ across all substrates). Under $U(1)_{B-L}$, however, it directly tracks the neutron fraction, which, for identical mirrors, varies by only a few percent among standard substrate candidates (silicon, fused silica, and sapphire).

Furthermore, incorporating an independent readout of the differential arm-length (DARM) mode to capture length degrees of freedom would provide key complementary constraints alongside the long-baseline strain channel for displacement-coupled ULDM candidates. This strategy directly builds upon the auxiliary-channel search frameworks pioneered by KAGRA~\cite{Michimura_2020, ULDM_Kagra_aux}. In particular, intentionally introducing a charge-to-mass ratio ($q_D/M$) mismatch between the test masses across the two arms breaks common-mode signal cancellation, thereby converting an otherwise suppressed common-mode response into a detectable differential strain signal. An analogous complementary search channel for dilaton DM leverages the scalar-induced modulation of optical properties within the circulating field inside a reference cavity~\cite{HA_2022}. Quantitative modeling of this channel will require detailed specifications of mirror coating thicknesses, refractive indices, and thermo-optic coefficients, which will become accessible as the mission's optical design matures.

In summary, this study highlights the compelling potential of the \indD detector for the direct detection of ULDM candidates. By demonstrating how a decihertz interferometer can simultaneously advance compact binary astrophysics while bridging critical parameter gaps in ULDM particle physics, our findings provide a strong rationale for integrating DM search strategies into the early architectural design of future space-based GW observatories. Looking ahead, dedicated instrumental trade studies will be essential—most notably to identify optimal material and subsystem configurations that maximize sensitivity to DM couplings without compromising \indD's primary GW science capabilities.

\acknowledgements 

We thank an anonymous referee for reviewing this paper. This work is partly supported by the SERB Start-up Research Grant No. SRG/2020/001290 funded by the Department of Science \& Technology (DST), Govt. of India, and Grant No. 12347103, 12547104 and W2611007 by the National Natural Science Foundation of China (NSFC), and Fundamental Research Funds for the Central Universities. A.M. acknowledges Rajesh K. Nayak for general encouragements in publishing this study. The authors also acknowledge Huaike Guo and Yue Zhao for useful discussions. 

\bibliography{references}

\end{document}